\documentclass[pra,twocolumn,amsmath,amssymb,superscriptaddress]{revtex4-1}
\usepackage{epsfig,amsmath}
\usepackage{subfigure}
\usepackage{graphicx}
\usepackage{dcolumn}
\usepackage{stmaryrd}
\usepackage{mathrsfs}
\usepackage{pifont}
\usepackage{amsthm}
\usepackage{amssymb}
\usepackage{bm}
\usepackage{latexsym}
\usepackage[colorlinks=true,linkcolor=blue,citecolor=blue]{hyperref}
\usepackage{color}
\usepackage{epstopdf}

\usepackage{booktabs}
\usepackage{threeparttable}
\usepackage{multirow}

\begin{document}

\title{Universal perspective on nonadiabatic quantum control}

\author{Zhu-yao Jin}
\affiliation{School of Physics, Zhejiang University, Hangzhou 310027, Zhejiang, China}

\author{Jun Jing}
\email{Email address: jingjun@zju.edu.cn}
\affiliation{School of Physics, Zhejiang University, Hangzhou 310027, Zhejiang, China}

\date{\today}

\begin{abstract}
A stable and fast path linking two arbitrary states of a quantum system is generally required for state-engineering protocols, such as stimulated Raman adiabatic passage, shortcuts to adiabaticity, and holonomic transformation. Such a path is also fundamental to the exact solution of the time-dependent Schr\"odinger equation. We construct a universal control framework using an ancillary picture, in which the time-dependent Hamiltonian can be diagonalized. Multiple desired paths can be derived by the von Neumann equation for parametric ancillary projection operators. No transition exists among the ancillary basis states during the time evolution. Under various conditions, our control framework reduces to the nonadiabatic holonomic transformation, the Lewis-Riesenfeld theory for invariants, and counterdiabatic driving methods. In addition, it is applicable to the cyclic transfer of system populations that could be a hard problem for existing methods. Our work can provide a full-rank time-evolution operator for a time-dependent quantum system with a finite number of dimensions.
\end{abstract}

\maketitle

\section{Introduction}

Real-time state control~\cite{Kral2007Colloquium} over quantum systems, such as holonomic transformation~\cite{Zanardi1999Holonomic} and coherent population transfer~\cite{Bergmann1998Coherent}, is fundamental and central to quantum information processing~\cite{Gisin2007Quantum,Kimble2008Quantum} and quantum networks~\cite{Kimble2008Quantum}. Holonomic transformation aims to manipulate the computational subspace for a desired geometric quantum computation. Coherent population transfer aims to transfer populations among discrete states with no dissipation. Intuitively, they seem distinct. The current work, however, attempts to find a systematic approach to understanding how they can be carried out under the same theoretical framework.

The geometric phase~\cite{Dong2021Doubly} is the core concept for geometric quantum gates, starting from the Abelian case~\cite{Berry1984Quantal,Anharonov1987Phase} in both adiabatic and nonadiabatic passages and extending to the non-Abelian case~\cite{Wilczek1984Appearance,Anandan1988Non,Zanardi1999Holonomic,De2003Berry,Leek2007Observation,Filipp2009Experimental} in the degenerate subspace. In holonomic transformation, the geometric phase provides an extra degree of freedom for the manipulation of computational states. The holonomy can be exemplified through parallel transport along closed cycles to preserve the geometrical information to be transported~\cite{Cohen2019Geometric}. Holonomic transformation is fault tolerant to local noise during the cyclic evolution, through which the geometric phase is accumulated by adiabatic~\cite{Born1928Beweis,Jones2000Geometric,Duan2001Geometric} or nonadiabatic~\cite{Sjoqvist2012Nonadiabatic,Liu2019Plug} driving. A system with an adiabatically changing Hamiltonian has no dynamical phase~\cite{Berry1984Quantal,Wilczek1984Appearance}, but suffers from the long exposure to the environment. In the nonadiabatic case, the dynamical phase remains vanishing at every moment under the parallel-transport condition~\cite{Sjoqvist2012Nonadiabatic}. But the system is sensitive to the errors in the parametric control~\cite{Zheng2016Comparison,Jing2017Non}. Through the spin-echo technique~\cite{Jones2000Geometric,Liu2019Plug} or the pulse-shaping method~\cite{Liu2019Plug}, the accumulation of the dynamical phase can be periodically eliminated during the time evolution. This means that the parallel-transport condition can be relaxed by path optimization~\cite{Liu2019Plug}.

Based on the adiabatic theorem~\cite{Claridge2009high,Vitanov2017Stiumlated}, a pedagogical model of the stimulated Raman adiabatic passage (STIRAP)~\cite{Vitanov2017Stiumlated} can be used to demonstrate coherent population transfer. It is frequently witnessed in a variety of disciplines, including atomic, molecular, and optical physics~\cite{Pillet1993Adiabatic,Phillips1998Nobel} and quantum information processing~\cite{Garc2003Quantum,Daems2007Adiabatic}. In comparison to resonant driving and Rabi oscillation, STIRAP is notably immune to the spontaneous emission from the intermediate state and robust against the fluctuation in experimental parameters. To avoid the non-negligible decoherence due to the longtime evolution of the open quantum system~\cite{Jing2016Eigenstate}, techniques inspired by the Lewis-Riesenfeld (LR) theory for invariants and counterdiabatic driving (CD) have been proposed under the name of shortcuts to adiabaticity (STA)~\cite{Chen2010Shortcut,Guery2019Shortcuts,Baksic2016Speeding,Li2016Shortcut,Chen2010Fast,Chen2011Lewis,
Chen2012Engineering,Qi2022Accelerated,An2016Shortcuts}. In particular, the LR-invariant method~\cite{Chen2011Lewis,Chen2012Engineering,Guery2019Shortcuts,Qi2022Accelerated} is applied to quantum systems with SU(2) dynamical symmetry, such as two-level atoms~\cite{Chen2011Lewis}, $\Lambda$-type qutrits~\cite{Chen2010Fast,Chen2012Engineering}, and continuous-variable systems with a quadratic Hamiltonian~\cite{Qi2022Accelerated}. Under transitionless quantum driving~\cite{Berry2009Transition}, the CD method is developed on the adiabatic path~\cite{Chen2010Shortcut,Guery2019Shortcuts,Qi2022Accelerated} and the dressed states~\cite{Baksic2016Speeding,Li2016Shortcut}. However, these protocols are reverse engineered~\cite{Guery2019Shortcuts}. Also, they have unnecessary complexities in motivation and are constrained in various scenarios.

We provide a universal framework of forward engineering to control the system dynamics. The time-dependent Hamiltonian can be fully or partially diagonalized within an ancillary picture. The solution to the von Neumann equation for the undetermined ancillary basis state yields a transitionless path for either shortcut-to-adiabaticity or holonomic transformation. When a full-rank evolution operator is constructed, i.e., all the ancillary basis states satisfy the von Neumann equation, the time-dependent Schr\"odinger equation can be exactly solved. Our universal framework simplifies the state-transfer protocols and unravels the true nature of holonomic transformation, STIRAP, STA, and reverse engineering. In this framework, we propose a protocol for cyclic population transfer, a hard or complex problem for existing methods.

The rest of this paper is structured as follows. In Sec.~\ref{general}, we introduce a general theoretical framework in which multiple nonadiabatic paths can be engineered under a time-dependent system Hamiltonian. Section~\ref{Secouruni} illustrates the general framework with a popular three-level system. A series of well-known protocols are then reproduced under the framework, including nonadiabatic holonomic transformation (Sec.~\ref{Secholonomic}), the Lewis-Riesenfeld theory for invariants (Sec.~\ref{SecLR}), and counterdiabatic driving methods (Secs.~\ref{SecCDd} and \ref{SecCDa}). Section~\ref{comparisonSec} delivers the main message of our paper: All these protocols can be unified with no unnecessary confusion. In Sec.~\ref{statetransfer}, we apply our universal protocol to cyclic population transfer, exemplified by a three-level system of the $\Lambda$ type (Sec.~\ref{threelevelsystem}) and a four-level system (Sec.~\ref{fourlevelsystem}). In Sec.~\ref{discussion}, we discuss the common and distinct points between our universal protocol and the others and provide a brief recipe for our protocol. The whole work is concluded in Sec.~\ref{conclusion}.

\section{General framework}\label{general}

Consider a quantum system of $N$ dimensions driven by a time-dependent Hamiltonian $H(t)$. In principle, we have a complete and orthonormal set of basis states $|\psi_m(t)\rangle$'s to span the whole Hilbert space of the system. Every $|\psi_m(t)\rangle$ is a pure-state solution to the Schr\"odinger equation ($\hbar\equiv1$):
\begin{equation}\label{Sch}
i\frac{d|\psi_m(t)\rangle}{dt}=H(t)|\psi_m(t)\rangle.
\end{equation}
$|\psi_m(0)\rangle$, $1\leq m\leq N$, represents the initial condition to be determined. Alternatively, the system dynamics can be described in an instantaneous picture with the basis states $|\mu_k(t)\rangle$, $1\leq k\leq N$, that span the same system Hilbert space. Distinct from $|\psi_m(t)\rangle$, the formation of the ancillary basis state $|\mu_k(t)\rangle$ is not directly determined by the system Hamiltonian. With unitary transformation, the solution picture $\{|\psi_m(t)\rangle\}$ and the ancillary picture $\{|\mu_k(t)\rangle\}$ can be connected as
\begin{equation}\label{transformation}
|\psi_m(t)\rangle=\sum_{k=1}^Nc_{mk}(t)|\mu_k(t)\rangle,
\end{equation}
where $c_{mk}(t)$ is the element of an $N\times N$ transformation matrix $\mathcal{C}$ in the $m$th row and the $k$th column. Substituting Eq.~(\ref{transformation}) into Eq.~(\ref{Sch}), we have collectively $N^2$ differential equations for all the matrix elements~\cite{Liu2019Plug}, i.e.,
\begin{equation}\label{element}
\frac{d}{dt}c_{mk}(t)=i\sum_{n=1}^N\left[\mathcal{G}_{kn}(t)-\mathcal{D}_{kn}(t)\right]c_{mn}(t),
\end{equation}
where $\mathcal{G}_{kn}(t)\equiv i\langle\mu_k(t)|\dot{\mu}_n(t)\rangle$ and  $\mathcal{D}_{kn}(t)\equiv\langle\mu_k(t)|H(t)|\mu_n(t)\rangle$ represent the geometric (not explicitly dependent on the system Hamiltonian) and dynamical (explicitly dependent on Hamiltonian) contributions to the integrating factor, respectively. It is, however, hard to directly solve Eq.~(\ref{element}) since every $c_{mk}(t)$ is coupled to the $N-1$ other elements $c_{mn}(t)$, $1\leq n\neq k\leq N$.

To simplify Eq.~(\ref{element}), one can consider the system Hamiltonian and the integrating factors in a time-independent ancillary picture $\{|\mu_k(0)\rangle\}$. With the unitary rotation by $V(t)\equiv\sum_{k=1}^N|\mu_k(t)\rangle\langle\mu_k(0)|$, we have
\begin{equation}\label{Hamrot}
\begin{aligned}
H_{\rm rot}(t)&=V^\dagger(t)H(t)V(t)-iV^\dagger(t)\frac{d}{dt}V(t)\\
&=-\sum_{k=1}^N\sum_{n=1}^N\left[\mathcal{G}_{kn}(t)-\mathcal{D}_{kn}(t)\right]|\mu_k(0)\rangle\langle\mu_n(0)|.
\end{aligned}
\end{equation}
It is found that {\em if} $H_{\rm rot}(t)$ is diagonal for certain $k$'s, $\{k\}\in\{1,2,\cdots,N\}$, then the integrating factor $\mathcal{G}_{kn}(t)-\mathcal{D}_{kn}(t)$ becomes vanishing unless $n=k$ and vice versa. The diagonalization could be full or partial. If and only if $\{k\}=\{1,2,\cdots,N\}$ is $H_{\rm rot}(t)$ fully diagonalized. In practice, one can regard $H_{\rm rot}(t)$ as its diagonalized part within a subspace spanned by eligible $|\mu_k(0)\rangle$'s, for which Eq.~(\ref{element}) is simplified to
\begin{equation}\label{digelement}
\frac{d}{dt}c_{mk}(t)=i\left[\mathcal{G}_{kk}(t)-\mathcal{D}_{kk}(t)\right]c_{mk}(t).
\end{equation}
It leads straightforwardly to the solution
\begin{equation}\label{cmk}
\begin{aligned}
c_{mk}(t)&=e^{if_k(t)}c_{mk}(0) \\
f_k(t)&\equiv\int_0^t\left[\mathcal{G}_{kk}(t')-\mathcal{D}_{kk}(t')\right]dt'.
\end{aligned}
\end{equation}
Due to the essential role played by the Schr\"odinger equation (\ref{Sch}), this solution is implicitly subject to the system Hamiltonian $H(t)$.

{\em Main result.} Here we prove that the von Neumann equation for the ancillary projection operator $\Pi_k(t)\equiv|\mu_k(t)\rangle\langle\mu_k(t)|$, $1\leq k\leq N$, e.g.,
\begin{equation}\label{von}
\frac{d}{dt}\Pi_k(t)=-i\left[H(t), \Pi_k(t)\right],
\end{equation}
is a necessary and sufficient condition for the diagonalization of $H_{\rm rot}(t)$ with a non vanishing diagonal element in the basis state $|\mu_k(0)\rangle$.

{\em Necessary condition.} When $H_{\rm rot}(t)$ is diagonal in the picture of $\{|\mu_k(0)\rangle\}$ and $\langle\mu_k(0)|H_{\rm rot}(t)|\mu_k(0)\rangle\ne0$, equivalently, we have the commutation expression 
\begin{equation}\label{notime}
\frac{d}{dt}\Pi_k(0)=-i\left[H_{\rm rot}(t), \Pi_k(0)\right]=0.
\end{equation}
If the diagonalization is not full-rank, then we can focus on the diagonalized subspace. With $\Pi_k(0)=V^\dagger\Pi_k(t)V$ and Eq.~(\ref{Hamrot}), Eq.~(\ref{notime}) can be recast as
\begin{equation}\label{vontime}
\frac{d\left[V^\dagger\Pi_k(t)V\right]}{dt}=-i\left[V^\dagger H(t)V-iV^\dagger\frac{d}{dt}V, V^\dagger\Pi_k(t)V\right].
\end{equation}
By the chain rule, it becomes
\begin{equation}\label{vontimeTrans}
\begin{aligned}
V^\dagger\frac{d\Pi_k(t)}{dt}V&=-i\left[V^\dagger H(t)V-iV^\dagger\frac{d}{dt}V, V^\dagger\Pi_k(t)V\right]\\
&-\frac{dV^\dagger}{dt}\Pi_k(t)V-V^\dagger\Pi_k(t)\frac{dV}{dt}.
\end{aligned}
\end{equation}
Multiplying Eq.~(\ref{vontimeTrans}) by $V(t)$ from the left and by $V^\dagger(t)$ from the right and using the complementary relation $(dV/dt)V^\dagger=-V(dV^\dagger/dt)$, we have
\begin{equation}\label{vontimeTransV}
\begin{aligned}
\frac{d\Pi_k(t)}{dt}&=-iV\left[V^\dagger H(t)V-iV^\dagger\frac{d}{dt}V, V^\dagger\Pi_k(t)V\right]V^\dagger\\
&-V\frac{dV^\dagger}{dt}\Pi_k(t)-\Pi_k(t)\frac{dV}{dt}V^\dagger\\
&=-i\left[H(t), \Pi_k(t)\right].
\end{aligned}
\end{equation}
Thus, the von Neumann equation~(\ref{von}) is a necessary condition for the diagonalization of $H_{\rm rot}(t)$ with nonvanishing elements $\langle\mu_k(0)|H_{\rm rot}(t)|\mu_k(0)\rangle$.

\emph{Sufficient condition.} The von Neumann equation~(\ref{von}) is equivalent to
\begin{equation}\label{idvon}
V^\dagger\frac{d\Pi_k(t)}{dt}V=-i\left[V^\dagger H(t)V, V^\dagger\Pi_k(t)V\right].
\end{equation}
By the chain rule, we have
\begin{equation}\label{commun}
\begin{aligned}
&\frac{d\left[V^\dagger\Pi_k(t)V\right]}{dt}=-i\left[V^\dagger H(t)V, V^\dagger\Pi_k(t)V\right]\\
+&\frac{dV^\dagger}{dt}\Pi_k(t)V+V^\dagger\Pi_k(t)\frac{dV}{dt}\\
=&-i\left[V^\dagger H(t)V-iV^\dagger\frac{dV}{dt}, V^\dagger\Pi_k(t)V\right].
\end{aligned}
\end{equation}
With the rotation back to the time-independent picture, it returns to Eq.~(\ref{notime}). Then we complete the proof for the sufficient condition for the diagonalization of $H_{\rm rot}(t)$ in the basis states $\{|\mu_k(0)\rangle\}$. A close justification of the sufficient condition can be found in Ref.~\cite{Liu2019Plug}, even though it focuses merely on a single path that can be solved by Eq.~(\ref{digelement}).

If all the ancillary basis states $|\mu_k(t)\rangle$, $1\leq k\leq N$, satisfy the von Neumann equation~(\ref{von}), then $H_{\rm rot}(t)$ is fully diagonal in the whole Hilbert space. Under such a strong condition, the exact solution for the time-dependent Schr\"odinger equation under $H(t)$ can be determined by the following time-evolution operator:
\begin{equation}\label{U}
U(t,0)=\sum_{k=1}^Ne^{if_k(t)}|\mu_k(t)\rangle\langle\mu_k(0)|,
\end{equation}
where $f_k(t)$ is the accumulated phase for the ancillary basis state $|\mu_k(t)\rangle$ in Eq.~(\ref{cmk}). In $U(t,0)$ no transition exists among all the ancillary states, so any of them can be an independent adiabatic path for (either accelerated or slowed) state or population transfer. In a deficient-rank case, $H_{\rm rot}(t)$ is diagonal with respect to $V(t)$ in a subspace spanned by $\tilde{N}<N$ ancillary basis states that satisfy Eq.~(\ref{von}). Then with proper initial conditions and reordering of the ancillary states, the system dynamics can be described by a reduced evolution operator, i.e., $\tilde{U}(t,0)=\sum_{k=1}^{\tilde{N}}e^{if_k(t)}|\mu_k(t)\rangle\langle\mu_k(0)|$.

The evolution operator $U(t,0)$ in Eq.~(\ref{U}) and its incomplete version $\tilde{U}(t,0)$ capture the basics of holonomic transformation, STA, and STIRAP. The phase $f_k(t)$ in Eqs.~(\ref{cmk}) or (\ref{U}) is not concerned with population transfer in STA and STIRAP that connects the initial state and the target state~\cite{Guery2019Shortcuts}. It can be an extra degree of freedom to manipulate the computational states in holonomic transformation. Many control methods such as STA can be categorized as reverse engineering. They determine the system Hamiltonian through the ansatz of a target evolution operator~\cite{Guery2019Shortcuts}. In contrast, our control framework provides a forward-engineering perspective that diagonalizes the given system Hamiltonian by performing a rotation $V(t)$ and gives rise to the time-evolution operator $U(t,0)$ with the von Neumann equation~(\ref{von}) for the ancillary projection operator. The unitary rotation $V(t)$ can be regarded as a quantum mechanical counterpart to the point transformation that transforms the ordinary coordinates of the system under constraint to the generalized coordinates in analytical mechanics. The output evolution determined by $V(t)$ in relation to the realistic evolution of the system $U(t,0)$ is analogous to the visual displacement in relation to the realistic displacement in classical mechanics.

\section{Our protocol versus existing control protocols}\label{comparisonSec}

\begin{figure}[htbp]
\centering
\includegraphics[width=0.7\linewidth]{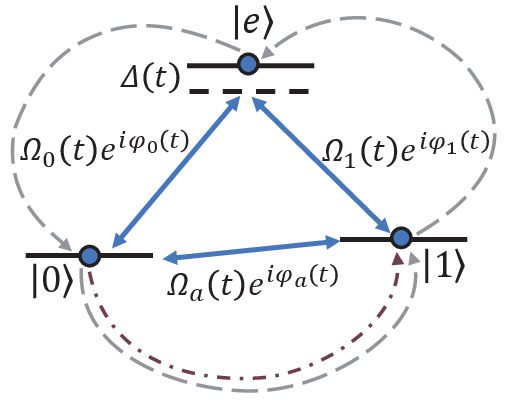}
\caption{Sketch of a three-level system under control, in which the transitions $|0\rangle\leftrightarrow|e\rangle$ and $|1\rangle\leftrightarrow|e\rangle$ are driven by fields of the same detuning and $|0\rangle\leftrightarrow|1\rangle$ is driven by a resonant driving field. The population transfer $|0\rangle\rightarrow|1\rangle$ and the cyclic population transfer $|0\rangle\rightarrow|1\rangle\rightarrow|e\rangle\rightarrow|0\rangle$ are described by the brown dot-dashed line and the gray dashed line, respectively.}\label{model}
\end{figure}

One can digest the utility and versatility of our framework by comparing it with well-established protocols for state control or population transfer~\cite{Sjoqvist2012Nonadiabatic,Chen2010Shortcut,Chen2012Engineering,Baksic2016Speeding,Liu2019Plug}. By modifying the parametric setting and boundary conditions, our general framework can be reduced to the nonadiabatic holonomic quantum transformation (NHQT)~\cite{Sjoqvist2012Nonadiabatic,Liu2019Plug}, the LR-invariant approach~\cite{Chen2012Engineering}, the CD method~\cite{Chen2010Shortcut}, and the method of counterdiabatic driving in the dressed states (CDD)~\cite{Baksic2016Speeding}. To incorporate these known protocols with a simplified discussion, we here focus on the population transfer from $|0\rangle$ to $|1\rangle$ in a three-level system of the $\Lambda$ type under control, which is marked by the brown dot-dashed line in Fig.~\ref{model}. Despite the theoretical and experimental efforts that have been made for decades, this pedagogical model contains all necessary elements of the problem at hand and sets our findings on a clear motivation.

\subsection{An illustrative example}\label{Secouruni}

Our protocol for population transfer in a three-level system (see Fig.~\ref{model}) is based on three classical driving fields. In particular, the transitions $|0\rangle\leftrightarrow|e\rangle$ and $|1\rangle\leftrightarrow|e\rangle$ are respectively driven by the laser fields $\Omega_0(t)$ and $\Omega_1(t)$ with the same detuning $\Delta(t)$. $\varphi_0(t)$ and $\varphi_1(t)$ are time-dependent phases. The transition between states $|0\rangle$ and $|1\rangle$ is driven by a resonant field with Rabi frequency $\Omega_a(t)$ and phase $\varphi_a(t)$. In experiments~\cite{Koch2007Charge,Antti2019Superadiabatic}, the transition $|0\rangle\leftrightarrow|1\rangle$ of the superconducting transmon qubit can be achieved with a two-photon process~\cite{Antti2019Superadiabatic}. Then the full Hamiltonian reads
\begin{equation}\label{Ham}
\begin{aligned}
&H(t)=\Delta(t)|e\rangle\langle e|+\Big[\Omega_0(t)e^{i\varphi_0(t)}|0\rangle\langle e|\\ +&\Omega_1(t)e^{i\varphi_1(t)}|1\rangle\langle e|+\Omega_a(t)e^{i\varphi_a(t)}|0\rangle\langle1|+{\rm H.c.}\Big].
\end{aligned}
\end{equation}

The system dynamics can be generally described in the ancillary picture spanned by the orthonormal time-dependent states $\{|\mu_0(t)\rangle, |\mu_1(t)\rangle, |\mu_2(t)\rangle\}$. They read
\begin{equation}\label{auxbasis}
\begin{aligned}
|\mu_0(t)\rangle&=\cos\theta(t)|0\rangle-\sin\theta(t)e^{-i\alpha(t)}|1\rangle, \\
|\mu_1(t)\rangle&=\sin\phi(t)|b(t)\rangle+\cos\phi(t)e^{-i\beta(t)}|e\rangle, \\
|\mu_2(t)\rangle&=\cos\phi(t)|b(t)\rangle-\sin\phi(t)e^{-i\beta(t)}|e\rangle,
\end{aligned}
\end{equation}
where $\theta(t)$, $\phi(t)$, $\alpha(t)$, and $\beta(t)$ are time-dependent parameters. $|b(t)\rangle\equiv\sin\theta(t)|0\rangle+\cos\theta(t)e^{-i\alpha(t)}|1\rangle$ and $|\mu_0(t)\rangle$ constitute a complete and orthonormal set of basis states in the subspace $\{|0\rangle, |1\rangle\}$. In the whole space, the ancillary basis states $|\mu_1(t)\rangle$ and $|\mu_2(t)\rangle$ are the superpositions over $|b(t)\rangle$ and the extra level $|e\rangle$, so that $|\mu_k(t)\rangle$, $k=0,1,2$, are orthonormal to each other. Equation~(\ref{auxbasis}) provides an explicit and parametric solution to construct the general time-evolution operator in Eq.~(\ref{U}). It is powerful enough to cover many protocols since the ancillary states $|\mu_k(t)\rangle$'s are capable of realizing the desired population transfer and accumulating the local phases of the discrete states.

\begin{figure}[htbp]
\centering
\includegraphics[width=0.9\linewidth]{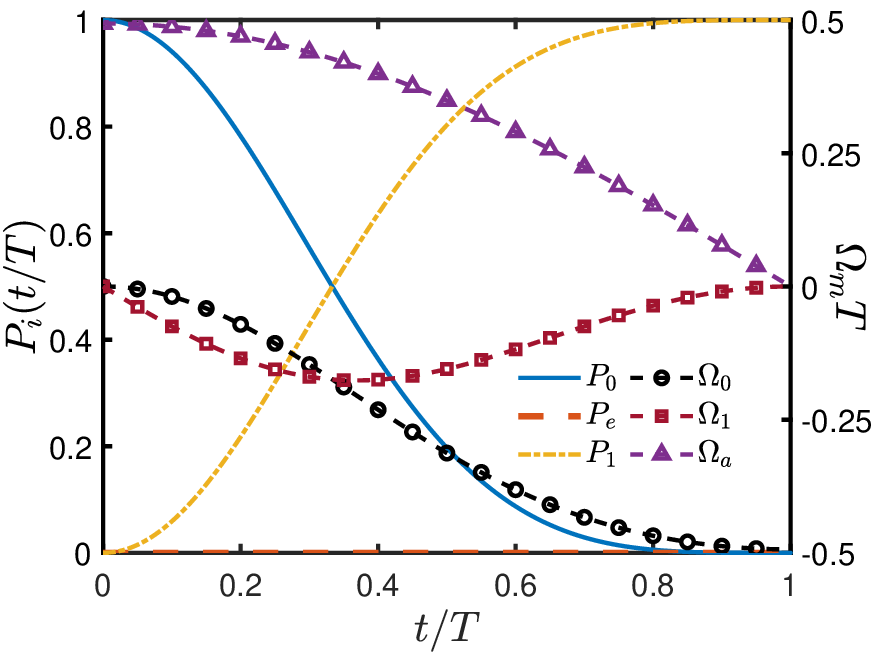}
\caption{State populations $P_i(t)$, $i=0,1,e$, and the driving frequencies $\Omega_m$, $m=0,1,a$, versus the evolution time $t/T$. The parameters are set according to $\varphi_i(t)$, $i=0,1,a$, in Eq.~(\ref{conditionphi}), $\Delta(t)$ in Eq.~(\ref{conditionDelta}), and $\Omega_i(t)$ in Eq.~(\ref{condition}). With $\varphi_0=\pi/2$, $\alpha=\beta=0$, $\theta(t)=(\pi/2)\sin[\pi t/(2T)]$, and $\phi(t)=(\pi/2)\cos[\pi t/(2T)]$, $\Delta(t)$ is found to be zero.}\label{universe}
\end{figure}

Substituting Eq.~(\ref{auxbasis}) into Eq.~(\ref{von}), we find that the phases are
\begin{equation}\label{conditionphi}
\begin{aligned}
 \varphi_0(t)-\varphi_1(t)&=\alpha(t),\\
 \varphi_a(t)-\alpha(t)&=-\arctan\left[\frac{2\dot{\theta}(t)}{\dot{\alpha}(t)\tan2\theta(t)}
 \right],
\end{aligned}
\end{equation}
the detuning is
\begin{equation}\label{conditionDelta}
\begin{aligned}
\Delta(t)&=2\Omega(t)\cot2\phi(t)\cos\left[\varphi_0(t)-\beta(t)\right]\\
&+\dot{\beta}(t)-\frac{1}{2}\dot{\alpha}(t)\tan2\theta(t)\cot\theta(t),\\
\end{aligned}
\end{equation}
and the Rabi frequencies are
\begin{equation}\label{condition}
\begin{aligned}
\Omega_0(t)&=\Omega(t)\sin\theta(t),\\
\Omega_1(t)&=\Omega(t)\cos\theta(t),\\
\Omega_a(t)&=\frac{\dot{\theta}(t)}{\sin\left[\varphi_a(t)-\alpha(t)\right]},
\end{aligned}
\end{equation}
where $\Omega(t)=\dot{\phi}(t)/\sin[\varphi_0(t)-\beta(t)]$. It is interesting to observe that the Hamiltonian~(\ref{Ham}) is full rank in the ancillary picture~(\ref{auxbasis}) since every $|\mu_k(t)\rangle$ can satisfy the von Neumann equation~(\ref{von}) under the same parametric setting in Eqs.~(\ref{conditionphi}), (\ref{conditionDelta}), and (\ref{condition}). Then any $|\mu_k(t)\rangle$ can be used as an accelerated adiabatic path. $\theta(t)$ and $\phi(t)$ and their boundary conditions determine the specific evolution from $|0\rangle$ to $|1\rangle$. In practice, the parameters $\theta(t)$ and $\phi(t)$ can be fully manipulated by the detuning $\Delta(t)$ and the driving intensities $\Omega_i(t)$, $i=0, 1, a$. When $\phi$ is set to be time independent, one cannot find proper conditions such as Eqs.~(\ref{conditionphi}), (\ref{conditionDelta}), and (\ref{condition}) to fully diagonalize the Hamiltonian in the ancillary picture $\{|\mu_k(t)\rangle\}$. In this case, the rank of the diagonalized part of $H_{\rm rot}(t)$ reduces to $1$. If one chooses $|\mu_0(t)\rangle$ for population transfer, then $\Omega(t)$ in Eqs.~(\ref{conditionDelta}) and (\ref{condition}) becomes a constant.

In the general case with time-dependent parameters, we use the path $|\mu_0(t)\rangle$ in Eq.~(\ref{auxbasis}) by setting the boundary conditions $\theta(0)=k\pi$ and $\theta(T)=\pi/2+k\pi$ with integer $k$. $\phi(t)$ is found to be irrelevant in this task. Using Eqs.~(\ref{conditionphi}), (\ref{conditionDelta}), and (\ref{condition}), the simulations of the state populations $P_i(t)=\langle i|\psi(t)\rangle\langle\psi(t)|i\rangle$, $i=0,1,e$, and the Rabi frequencies $\Omega_m(t)$, $m=0,1,a$, versus the evolution time $t$ are plotted in Fig.~\ref{universe}. Here the system dynamics is obtained with the time-dependent Schr\"odinger equation $i\partial |\psi(t)\rangle/\partial t=H(t)|\psi(t)\rangle$ with the Hamiltonian in Eq.~(\ref{Ham}) from the initial state $|\psi(0)\rangle=|0\rangle$. In particular, the detuning $\Delta$ is set to zero by letting $\varphi_0=\pi/2$ and $\alpha=\beta=0$, and the Rabi frequencies are parameterized as $\theta(t)=(\pi/2)\sin[\pi t/(2T)]$ and $\phi(t)=(\pi/2)\cos[\pi t/(2T)]$. It is found that the population on the state $|0\rangle$ can be completely and smoothly transferred to the state $|1\rangle$ after a period $T$. Along the path $|\mu_0(t)\rangle$, there is no occupation on the unwanted state $|e\rangle$.

\subsection{Nonadiabatic holonomic transformation}\label{Secholonomic}

Our universal protocol for the three-level system can be reduced to the nonadiabatic holonomic transformation~\cite{Sjoqvist2012Nonadiabatic,Liu2019Plug} when the external driving between $|0\rangle\leftrightarrow|1\rangle$ in Fig.~\ref{model} is turned off. Then with $\Delta(t)=0$ for simplicity, the full Hamiltonian in Eq.~(\ref{Ham}) becomes
\begin{equation}\label{HamGates}
H(t)=\frac{\Omega_0(t)}{2}e^{i\varphi_0(t)}|0\rangle\langle e|+\frac{\Omega_1(t)}{2}e^{i\varphi_1(t)}|1\rangle\langle e|+{\rm H.c.},
\end{equation}
where the time-dependent phases $\varphi_0(t)$ and $\varphi_1(t)$ contribute to the desired geometric phase. In this application, the excited state $|e\rangle$ serves as the ancillary state, and the lower states $|0\rangle$ and $|1\rangle$ constitute the computational subspace for quantum gates. In the dark-bright states for the computational subspace [see, e.g., Eq.~(\ref{Ugates})], it turns out that a local geometric phase accumulates on the bright state $|b(t)\rangle$ as an extra degree of freedom determining the quantum gate.

Practically, the ancillary picture can be spanned with the following orthonormal ancillary states:
\begin{equation}\label{auxbasisGates}
\begin{aligned}
|\mu_0(t)\rangle&=\cos\frac{\theta(t)}{2}|0\rangle+\sin\frac{\theta(t)}{2}e^{i\varphi(t)}|1\rangle,\\
|\mu_1(t)\rangle&=\sin\frac{\phi(t)}{2}|b(t)\rangle+\cos\frac{\phi(t)}{2}e^{-i\alpha(t)}|e\rangle,\\
|\mu_2(t)\rangle&=\cos\frac{\phi(t)}{2}|b(t)\rangle-\sin\frac{\phi(t)}{2}e^{-i\alpha(t)}|e\rangle,
\end{aligned}
\end{equation}
where $|b(t)\rangle=\sin[\theta(t)/2]|0\rangle-\cos[\theta(t)/2]e^{i\varphi(t)}|1\rangle$; $\varphi(t)\equiv\varphi_1(t)-\varphi_0(t)$; and $\theta(t)$, $\phi(t)$, and $\alpha(t)$ are generally time dependent.

One can find that Eq.~(\ref{auxbasisGates}) is a specific formation of Eq.~(\ref{auxbasis}) by following the convention in Refs.~\cite{Sjoqvist2012Nonadiabatic,Liu2019Plug}, i.e., resetting the parameters as $\theta(t)\rightarrow\theta(t)/2$, $\phi(t)\rightarrow\phi(t)/2$, $\alpha(t)\rightarrow-\varphi(t)+\pi$, and $\beta(t)\rightarrow\alpha(t)$. Substituting all three states of Eq.~(\ref{auxbasisGates}) into the von Neumann equation~(\ref{von}), we find that $\theta(t)$ is no longer dependent on time, i.e.,
\begin{equation}\label{thetaHol}
\theta(t)\rightarrow\theta,
\end{equation}
and the Rabi frequencies and the phase are
\begin{equation}\label{RabiGates}
\begin{aligned}
\Omega_0(t)&=\Omega(t)\sin\frac{\theta}{2},\\
\Omega_1(t)&=-\Omega(t)\cos\frac{\theta}{2},\\
\varphi_0(t)&=\alpha(t)-\arctan\left[\frac{\dot{\phi}(t)}{\dot{\alpha}(t)\tan\phi(t)}\right],
\end{aligned}
\end{equation}
respectively, where $\Omega(t)=\dot{\phi}(t)/\sin[\varphi_0(t)-\alpha(t)]$. With Eqs.~(\ref{thetaHol}) and (\ref{RabiGates}), $|\mu_0(t)\rangle\rightarrow|\mu_0\rangle$ is found to be a static dark state with zero eigenvalue and is decoupled from the system dynamics. The system dynamics is then constrained in the subspace spanned by $\{|\mu_1(t)\rangle, |\mu_2(t)\rangle\}$. Our result is consistent with the conventional assumption~\cite{Sjoqvist2012Nonadiabatic,Liu2019Plug} that $\theta$ is time independent in the holonomic transformation.

Equation~(\ref{U}) indicates that if the system starts from the computational subspace and goes back to the same one through the path $|\mu_1(t)\rangle$ or $|\mu_2(t)\rangle$ (under respective boundary conditions), then a phase $f_k(t)$ can be generated on the bright state $|b(t)\rangle$ relative to the dark state $|\mu_0(t)\rangle$. That phase becomes purely geometric when the dynamical part is eliminated.

\begin{figure}[htbp]
\centering
\includegraphics[width=0.9\linewidth]{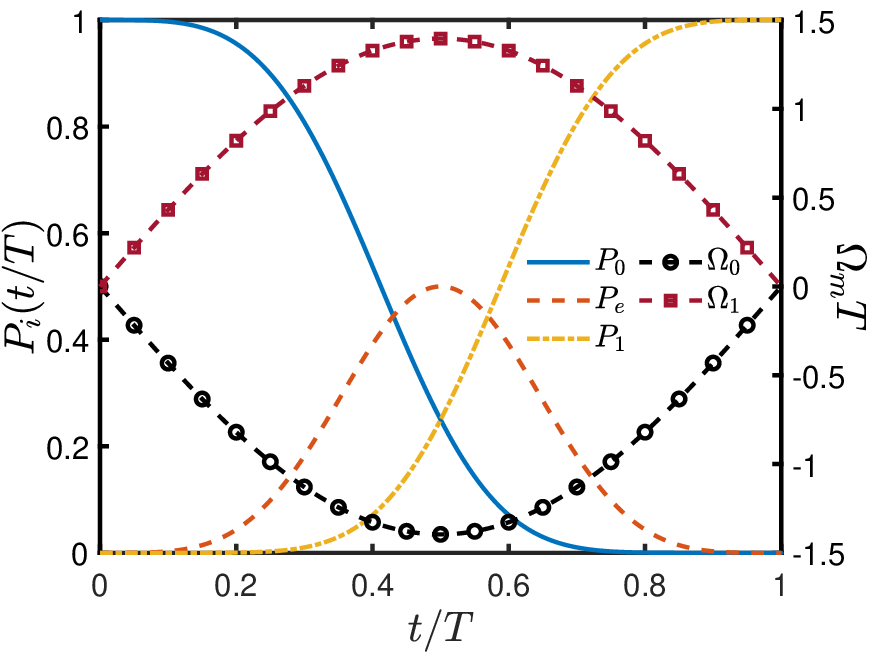}
\caption{State populations $P_i(t)$, $i=0,1,e$, and the driving frequencies $\Omega_m$, $m=0,1$, versus the evolution time $t/T$. The parameters are set according to $\Omega_0(t)$, $\Omega_1(t)$, and $\varphi_0(t)$ in Eq.~(\ref{RabiGates}) with $\phi(t)=\pi\cos(\pi t/T)$ and $\alpha(t)=2\pi\Theta(T/2)$.}\label{holonomic}
\end{figure}

In particular, if the system evolves along the path $|\mu_1(t)\rangle$, then the phase can be expressed as $f_1(t)=\gamma_g(t)+\gamma_d(t)$ using Eqs.~(\ref{digelement}) and (\ref{cmk}). It is a summation of the geometric phase $\gamma_g(t)=\int_0^t\mathcal{G}_{11}(t')dt'$, with $\mathcal{G}_{11}(t)=\dot{\alpha}(t)\cos^2[\phi(t)/2]$, and the dynamical phase $\gamma_d(t)=\int_0^t\mathcal{D}_{11}(t')dt'$, with $\mathcal{D}_{11}(t)=\Omega(t)\sin\phi(t)\cos[\varphi_0(t)-\alpha(t)]$. The dynamical phase remains zero when $\varphi_0(t)-\alpha(t)=\pi/2$, which is consistent with the parallel-transport condition~\cite{Sjoqvist2012Nonadiabatic}, or can vanish after a step wise cancellation~\cite{Jones2000Geometric,Liu2019Plug}. Under the parallel-transport condition, the geometric phase $\gamma_g$ can be simply obtained by setting $\alpha(t)$ or $\varphi_0(t)$ as a step function. For example, when $\alpha(t)=2\gamma\Theta(T/2)$ with the Heaviside step function $\Theta(t)$ and $\phi(t)=\pi\cos(\pi t/T)$ [with regard to the boundary condition of the path $|\mu_1(t)\rangle$], the geometric phase is found to be $\gamma_g=\gamma$. Then the evolution operator for the holonomic transformation can be written as
\begin{equation}\label{Ugates}
\begin{aligned}
U(\theta, \varphi(t), \gamma)&=|\mu_0(t)\rangle\langle\mu_0(t)|-e^{i\gamma}|b(t)\rangle\langle b(t)|\\
&=e^{i\gamma/2}e^{-i\gamma/2\vec{n}\cdot\vec{\sigma}},
\end{aligned}
\end{equation}
where $\vec{n}\equiv(\sin\theta\cos\varphi(t), -\sin\theta\sin\varphi(t), \cos\theta)$ is the rotating axis. According to Eq.~(\ref{RabiGates}), $\theta$ is determined by the intensity ratio of the two driving fields, and according to Eq.~(\ref{auxbasisGates}), $\varphi(t)$ is their phase difference. The population on $|0\rangle$ can be transferred to $|1\rangle$ by $U(\pi/2,0,\pi)$ in Eq.~(\ref{Ugates}), although this task is not the real intention of $U(\theta,\varphi(t),\gamma)$. In Fig.~\ref{holonomic}, we plot the population dynamics for $|i\rangle$, $i=0,1,e$, and the Rabi frequencies $\Omega_m(t)$, $m=0,1$, in Eq.~(\ref{RabiGates}). The initial population on the state $|0\rangle$ is found to be completely transferred to the target state $|1\rangle$ at the end of the period. In contrast to Fig.~\ref{universe}, with our protocol, the ancillary state $|e\rangle$ can be temporarily occupied due to the formalization of $|\mu_1(t)\rangle$, e.g., $P_e(t)=0.5$ when $t=T/2$. In addition to the population transfer, Eqs.~(\ref{auxbasisGates}) and (\ref{Ugates}) indicate that the phase $\gamma$ of the geometric gate can be manipulated by the phases of the computational states, such as $\alpha(t)$ and $\phi(t)$.

\subsection{Lewis-Riesenfeld theory for invariants}\label{SecLR}

In the control field of the shortcut to adiabaticity, the LR theory for invariants is popular for systems with SU(2) dynamical symmetry, including discrete and continuous-variable systems. In Fig.~\ref{model}, when the driving between $|0\rangle$ and $|1\rangle$ is absent and the other two driving fields are resonant and in phase, the invariant theory~\cite{Chen2012Engineering} can emerge from our universal protocol. In this case, the full Hamiltonian in Eq.~(\ref{Ham}) can be reduced to
\begin{equation}\label{HamLewis}
H(t)=\Omega_0(t)|0\rangle\langle e|+\Omega_1(t)|1\rangle\langle e|+{\rm H.c.}.
\end{equation}

The system dynamics can be described in the ancillary picture with $\{|\mu_0(t)\rangle, |\mu_1(t)\rangle, |\mu_2(t)\rangle\}$ as
\begin{equation}\label{auxbasisLewis}
\begin{aligned}
|\mu_0(t)\rangle&=\sin{\theta(t)}|0\rangle-\cos{\theta(t)}|1\rangle,\\
|\mu_1(t)\rangle&=\cos{\theta(t)}\sin{\phi(t)}|0\rangle+i\cos{\phi(t)}|e\rangle\\
&+\sin{\theta(t)}\sin{\phi(t)}|1\rangle,\\
|\mu_2(t)\rangle&=\cos{\theta(t)}\cos{\phi(t)}|0\rangle-i\sin{\phi(t)}|e\rangle\\
&+\sin{\theta(t)}\cos{\phi(t)}|1\rangle
\end{aligned}
\end{equation}
with the time-dependent parameters $\theta(t)$ and $\phi(t)$. The states in Eq.~(\ref{auxbasisLewis}) can be obtained by the general case in Eq.~(\ref{auxbasis}) when $\theta(t)\rightarrow\pi/2-\theta(t)$, $\alpha(t)=0$, and $\beta(t)=-\pi/2$. Instead of demonstrating the system dynamics within the whole Hilbert space, here one can focus merely on the one-dimensional ancillary subspace based on $|\mu_1(t)\rangle$ or $|\mu_2(t)\rangle$, which is a superposed state of $|0\rangle$, $|1\rangle$, and $|e\rangle$. In other words, one cannot achieve a full-rank diagonalized Hamiltonian $H_{\rm rot}(t)$ through the rotation in Eq.~(\ref{Hamrot}) with the Hamiltonian~(\ref{HamLewis}) and the ancillary states in Eq.~(\ref{auxbasisLewis}). The von Neumann equation~(\ref{von}) fails to yield a consistent group of parametric settings unless we input only a single state, $|\mu_1(t)\rangle$ or $|\mu_2(t)\rangle$. With $|\mu_2(t)\rangle$, the Rabi frequencies $\Omega_0(t)$ and $\Omega_1(t)$ are found to take the same form as in Ref.~\cite{Chen2012Engineering}:
\begin{equation}\label{OmLewis}
\begin{aligned}
\Omega_0(t)&=\dot{\theta}(t)\sin\theta(t)\cot\phi(t)+\dot{\phi}(t)\cos\theta(t),\\
\Omega_1(t)&=-\dot{\theta}(t)\cos\theta(t)\cot\phi(t)+\dot{\phi}(t)\sin\theta(t).
\end{aligned}
\end{equation}
In regard to the cotangent function, the parameter $\phi(t)$ should be properly chosen to avoid the divergent behavior of the Rabi frequencies $\Omega_0(t)$ and $\Omega_1(t)$. Using Eq.~(\ref{U}), the relevant time-evolution operator can be written as
\begin{equation}\label{ULewis}
U(t,0)=e^{if_2(t)}|\mu_2(t)\rangle\langle\mu_2(0)|,
\end{equation}
where $f_2(t)$ is the Lewis-Riesenfeld phase~\cite{Guery2019Shortcuts} and it is not relevant to the population transfer.

\begin{figure}[htbp]
\centering
\includegraphics[width=0.9\linewidth]{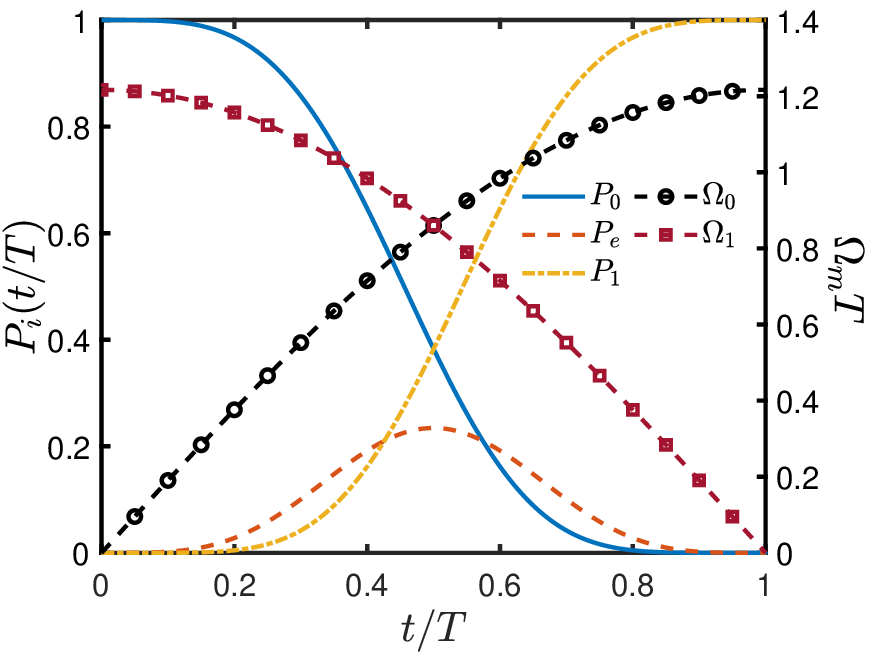}
\caption{State populations $P_i(t)$, $i=0,1,e$, and the driving frequencies $\Omega_m$, $m=0,1$, versus the evolution time $t/T$. The parameters are set according to $\Omega_0(t)$ and $\Omega_1(t)$ in Eq.~(\ref{OmLewis}) with $\theta(t)=\pi-\pi t/(2T)$ and $\phi=0.2527$~\cite{Chen2012Engineering}.}\label{Lewis}
\end{figure}

Using Eq.~(\ref{OmLewis}) with $\theta(t)=\pi-\pi t/(2T)$ and $\phi=0.2527$~\cite{Chen2012Engineering}, we demonstrate in Fig.~\ref{Lewis} the time evolution of populations $P_i(t)$, $i=0,1,e$, using the time-dependent Schr\"odinger equation as well as the relevant Rabi frequencies $\Omega_m(t)$, $m=0,1$. The three-level system is initially prepared as $|0\rangle$. The system population can be completely transferred to $|1\rangle$ after a period $T$. However, during this evolution process, the mediated state $|e\rangle$ can be temporally populated due to the formalism of $|\mu_2(t)\rangle$. $P_e(t)$ is as high as $0.22$ when $t=T/2$.

\subsection{Counterdiabatic driving in dressed states}\label{SecCDd}

Our control protocol can recover the CDD method~\cite{Baksic2016Speeding} when the driving field on $|0\rangle\leftrightarrow|1\rangle$ in Fig.~\ref{model} is turned off, the driving fields on $|0\rangle\leftrightarrow|e\rangle$ and $|1\rangle\leftrightarrow|e\rangle$ are resonant in frequency, and their phases are set to be time independent. Accordingly, the Hamiltonian in Eq.~(\ref{Ham}) can be written as
\begin{equation}\label{HamDress}
H(t)=\Omega_0(t)e^{i\varphi_0}|0\rangle\langle e|+\Omega_1(t)e^{i\varphi_1}|1\rangle\langle e|+{\rm H.c.}.
\end{equation}
Then the ancillary picture can be spanned by those states in Eq.~(\ref{auxbasis}) with $\alpha=\beta=0$. Using the components in this ancillary picture, the von Neumann equation~(\ref{von}) yields
\begin{equation}\label{PhaseDress}
\begin{aligned}
&\varphi_0=\varphi_1=\frac{\pi}{2},\\
&\theta(t)\rightarrow\theta,
\end{aligned}
\end{equation}
and
\begin{equation}\label{RabiDress}
\begin{aligned}
\Omega_0(t)&=\Omega(t)\sin\theta,\\
\Omega_1(t)&=\Omega(t)\cos\theta,
\end{aligned}
\end{equation}
where $\Omega(t)=\dot{\phi}(t)$. Under the conditions in Eqs.~(\ref{PhaseDress}) and (\ref{RabiDress}), $|\mu_0\rangle$ is time independent and decoupled from the system dynamics. It is found that Eq.~(\ref{RabiDress}) can be recovered either by Eq.~(\ref{RabiGates}) for the nonadiabatic holonomic transformation when $\varphi_0-\alpha=\pi/2$ and $\theta/2\rightarrow-\theta+\pi$ or by Eq.~(\ref{OmLewis}) for the Lewis-Riesenfeld theory when $\theta(t)$ is time independent. These relations are expectable since the three protocols stem from almost the same evolution path.

\begin{figure}[htbp]
\centering
\includegraphics[width=0.9\linewidth]{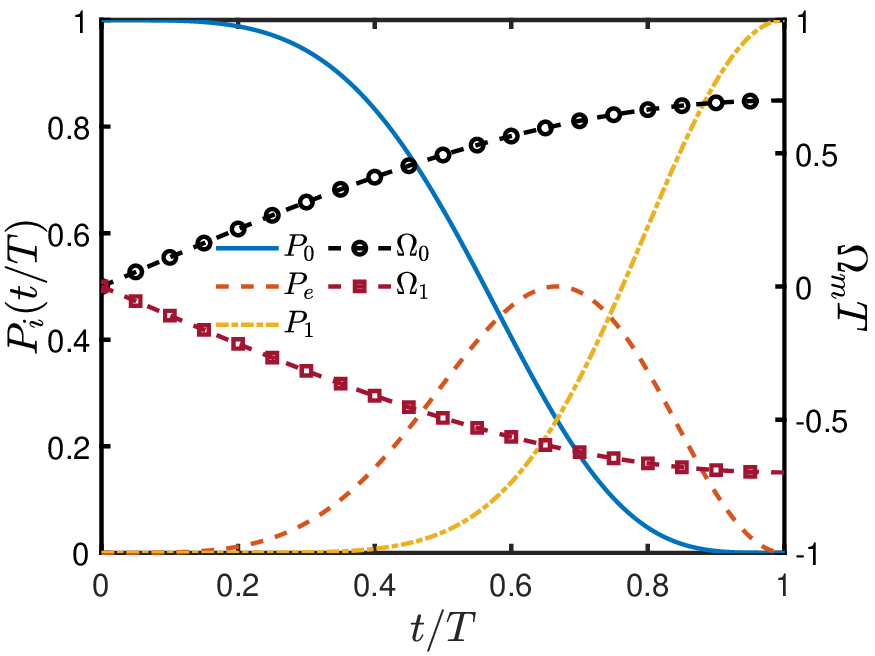}
\caption{State populations $P_i(t)$, $i=0,1,e$, and the driving frequencies $\Omega_m$, $m=0,1$, versus the evolution time $t/T$. The parameters are set according to $\Omega_0(t)$ and $\Omega_1(t)$ in Eq.~(\ref{RabiDress}) with $\theta=-\pi/4$ and $\phi(t)=\pi\cos[\pi t/(2T)]$.}\label{dressedstate}
\end{figure}

With no loss of generality, the system is supposed to evolve along the path $|\mu_2(t)\rangle$. Using Eq.~(\ref{RabiDress}) with $\theta=-\pi/4$ and $\phi(t)=\pi\cos[\pi t/(2T)]$, we demonstrate the population dynamics $P_i(t)$ and Rabi frequencies in Fig.~\ref{dressedstate}. Although the initial population on the state $|0\rangle$ can be fully transferred to the state $|1\rangle$ after a period $T$, the mediated state $|e\rangle$ can be significantly populated. When $t=0.65T$, $P_e(t)=0.5$. Distinct from Figs.~\ref{universe}, \ref{holonomic}, and \ref{Lewis}, the whole pattern is asymmetrical to $t=T/2$.

\subsection{Counterdiabatic driving with ancillary Hamiltonian}\label{SecCDa}

When the detuning $\Delta$ of the driving fields on the transitions $|0\rangle\leftrightarrow|e\rangle$ and $|1\rangle\leftrightarrow|e\rangle$ in Fig.~\ref{model} and the phases $\varphi_0$, $\varphi_1$, and $\varphi_a$ are time independent, our universal protocol can simulate the counterdiabatic driving method via adiabatic passage~\cite{Chen2010Shortcut}. In the language of quantum transitionless driving~\cite{Berry2009Transition}, the full Hamiltonian~(\ref{Ham}) can be decomposed to the reference Hamiltonian $H_0(t)$ and the ancillary Hamiltonian $H_c(t)$ as
\begin{equation}\label{HamCounterfull}
H(t)=H_0(t)+H_c(t),
\end{equation}
where
\begin{equation}\label{HamCounter}
\begin{aligned}
H_0(t)&=\Delta|e\rangle\langle e|+\Omega_0(t)e^{i\varphi_0}|0\rangle\langle e|+\Omega_1(t)e^{i\varphi_1}|1\rangle\langle e|,\\
H_c(t)&=\Omega_a(t)e^{i\varphi_a}|0\rangle\langle1|+{\rm H.c.}.
\end{aligned}
\end{equation}

The basic idea of the counterdiabatic driving method is that the system can follow one instantaneous eigenstate of $H_0(t)$ with the assistance of $H_c(t)$. We demonstrate that the transitionless driving algorithm~\cite{Berry2009Transition} can be described within the ancillary picture spanned by the ancillary states $|\mu_k(t)\rangle$, $k=0, 1, 2$, in Eq.~(\ref{auxbasis}) with $\alpha=\beta=0$. In this picture, the von Neumann equation~(\ref{von}) yields
\begin{equation}\label{conditionCounterPhase}
\varphi_a=\frac{\pi}{2}, \quad \varphi_0=\varphi_1=0,
\end{equation}
and the detuning and the Rabi frequencies are
\begin{equation}\label{conditionCounter}
\begin{aligned}
&\Delta=2\Omega(t)\cot2\phi(t),\\
&\Omega_0(t)=\Omega(t)\sin\theta(t),\\
&\Omega_1(t)=\Omega(t)\cos\theta(t),\\
&\Omega_a(t)=\dot{\theta}(t),
\end{aligned}
\end{equation}
where $\Omega(t)=\dot{\phi}(t)$ can be regarded as a field intensity for normalization.

In the case with a fixed $\phi$, both $\Delta$ and $\Omega$ turn out to be time independent. We then recover the counterdiabatic driving method for the adiabatic passage~\cite{Chen2010Shortcut}. Along the same evolution path $|\mu_0(t)\rangle$, our universal protocol yields the same results for the populations $P_i(t)$, $i=0,1,e$, as those in Ref.~\cite{Chen2010Shortcut}. They can be exactly described by Fig.~\ref{universe}. In the general case with $\phi\rightarrow\phi(t)$, our protocol yields a full-rank nonadiabatic evolution operator, and it can be used to transfer the system population to the state $|e\rangle$ along $|\mu_1(t)\rangle$ or $|\mu_2(t)\rangle$. In other words, the application range of the counterdiabatic driving method can be widely expanded in our universal framework.

\section{Cyclic population transfer}\label{statetransfer}

More than demystifying and unifying the existing protocols on quantum state engineering~\cite{Sjoqvist2012Nonadiabatic,Chen2010Shortcut,Chen2012Engineering,Baksic2016Speeding,Liu2019Plug}, our protocol in Sec.~\ref{general} is sufficiently broad to go beyond their reach. This section exemplifies the cyclic population transfers in a three-level system (see the gray dashed lines in Fig.~\ref{model}) and in a four-level system (see Fig.~\ref{modelfour} below). They demonstrate the control power of our protocol in the unidirectional state transfer.

\subsection{Three-level system}\label{threelevelsystem}

Continuous cyclic population transfer among all three levels, $|0\rangle$, $|e\rangle$, and $|1\rangle$, could be a repetition of a single loop of $|0\rangle\rightarrow|e\rangle\rightarrow|1\rangle\rightarrow|0\rangle$, with every loop starting from the system initial state $|0\rangle$. We can divide such a single loop into two stages.

In stage $1$, one can employ the (accelerated) adiabatic path
\begin{equation}\label{Stage1}
\begin{aligned}
|\mu_2(t)\rangle&=\sin\theta(t)\cos\phi(t)|0\rangle-\sin\phi(t)e^{-i\beta(t)}|e\rangle\\
&+\cos\theta(t)\cos\phi(t)e^{-i\alpha(t)}|1\rangle
\end{aligned}
\end{equation}
in Eq.~(\ref{auxbasis}), which is an ansatz for one of the ancillary states with nonvanishing population on $|e\rangle$. The three ancillary basis states support a time-dependent picture for the system with a general Hamiltonian in Eq.~(\ref{Ham}). The driving phases, detuning, and intensities in Eqs.~(\ref{conditionphi}), (\ref{conditionDelta}), and (\ref{condition}) can ensure that no transition exists among the ancillary basis states during the whole transfer process. In particular, the parameters in $|\mu_2(t)\rangle$ can be set as
\begin{equation}\label{boundu2}
\theta(t)=\frac{\pi}{2}\cos\left(\frac{\pi t}{2T}\right),\quad \phi(t)=\frac{\pi}{2}\sin\left(\frac{\pi t}{T}\right).
\end{equation}
Then the population is transferred from $|0\rangle$ to $|e\rangle$ at $t=T/2$ and then transferred to $|1\rangle$ at $t=T$.

In stage $2$, one can employ the path
\begin{equation}\label{Stage2}
|\mu_0(t)\rangle=\cos\theta(t)|0\rangle-\sin\theta(t)e^{-i\alpha(t)}|1\rangle
\end{equation}
in Eq.~(\ref{auxbasis}) to move the system population from $|1\rangle$ back to $|0\rangle$. In particular, we set
\begin{equation}\label{boundu0}
\theta(t)=\frac{\pi}{2}\cos\left(\frac{\pi t}{T}\right),\quad \phi(t)=\frac{\pi}{2}\sin\left(\frac{2\pi t}{T}\right)
\end{equation}
for the boundary conditions of the path $|\mu_0(t)\rangle$. This stage lasts $T/2$; then the first loop is completed.

In general, the $k$th loop of cyclic population transfer during $t\in[3(k-1)T/2, 3kT/2]$, $k\geq1$, can be divided into two stages. In the first one, the parameters in $|\mu_2(t)\rangle$ can be set as
\begin{equation}\label{boundu2second}
\begin{aligned}
\theta(t)&=\frac{\pi}{2}\cos\left[\frac{\pi(2t-(3k-1)T)}{4T}\right],\\
\phi(t)&=-\frac{\pi}{2}\cos\left[\frac{\pi(2t-3kT)}{2T}\right],
\end{aligned}
\end{equation}
then the population on $|0\rangle$ is transferred to $|e\rangle$ at $t=(3k-2)T/2$, and transferred to $|1\rangle$ at $t=(3k-1)T/2$. In the second one, via the path $|\mu_0(t)\rangle$, we have
\begin{equation}\label{boundu0second}
\begin{aligned}
\theta(t)&=-\frac{\pi}{2}\cos\left[\frac{\pi(2t-3kT)}{2T}\right],\\
\phi(t)&=\frac{\pi}{2}\sin\left[\frac{\pi(2t-3kT)}{T}\right],
\end{aligned}
\end{equation}
which transfers the population on $|1\rangle$ back to $|0\rangle$ at $t=3kT/2$.

\begin{figure}[htbp]
\centering
\includegraphics[width=0.9\linewidth]{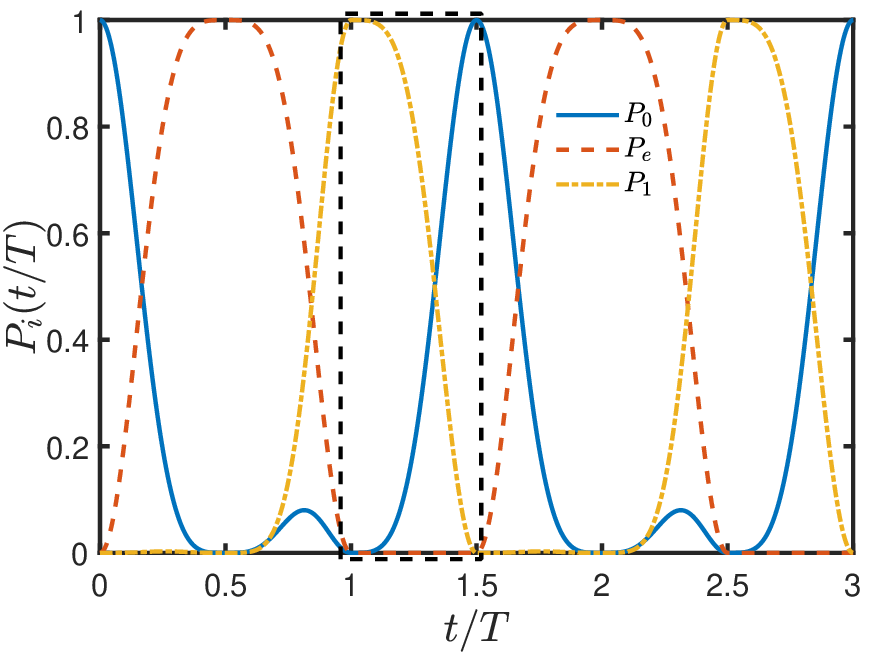}
\caption{State populations $P_i(t)$, $i=0,1,e$, versus the evolution time $t/T$ during the cyclic population transfer. The transfer process via the path $|\mu_0(t)\rangle$ in the first loop is marked with a black dashed frame. The phases $\varphi_i(t)$, $i=0,1,a$, are set according to Eq.~(\ref{conditionphi}) with $\alpha=0$ and $\varphi_0=\pi/2$. The Rabi frequencies $\Omega_i(t)$ are set according to Eq.~(\ref{condition}) with $\beta=0$, $\theta(t)$ and $\phi(t)$ in Eqs.~(\ref{boundu2}), (\ref{boundu0}), (\ref{boundu2second}), and (\ref{boundu0second}).}\label{transfer}
\end{figure}

Using Eqs.~(\ref{conditionphi}), (\ref{conditionDelta}), and (\ref{condition}) with $\alpha=\beta=0$ and $\theta(t)$ and $\phi(t)$ given in Eqs.~(\ref{boundu2}), (\ref{boundu0}), (\ref{boundu2second}), and (\ref{boundu0second}), we demonstrate in Fig.~\ref{transfer} the population dynamics $P_i(t)$ of the three levels in the cyclic transfer. It is shown that every loop of $|0\rangle\rightarrow|e\rangle\rightarrow|1\rangle\rightarrow|0\rangle$ is completed in a period of $3T/2$. The population transfers from $|0\rangle$ to $|e\rangle$ during $t\in[0, T/2]$ and from $|1\rangle$ to $|0\rangle$ during $t\in[T, 3T/2]$ are faithful since the respective unwanted states $|1\rangle$ and $|e\rangle$ are hardly populated. In contrast, during the transfer from $|e\rangle$ to $|1\rangle$, i.e., $t\in[T/2, T]$, the state $|0\rangle$ can be slightly populated, e.g., $P_0(t)=0.08$ when $t=0.82T$. However, it does not obstruct the complete transfer $P_1(t=T)=1$. In addition, the transfer process during $t\in[T, 3T/2]$, as outlined with the black dashed frame in Fig.~\ref{transfer}, is a typical evolution observed in both STIRAP~\cite{Kral2007Colloquium,Vitanov2017Stiumlated} and counterdiabatic driving~\cite{Chen2010Shortcut}. This process does not involve with the intermediate state $|e\rangle$ due to the popular path $|\mu_0(t)\rangle$ in Eq.~(\ref{Stage2}).

More generally, the population on any initial state can be transferred to any target state via any ancillary path if it is permitted by the boundary condition. Alternatively, the loop of cyclic transfer can be $|0\rangle\rightarrow|1\rangle\rightarrow|e\rangle\rightarrow|0\rangle$. For example, along the path $|\mu_0(t)\rangle$ with $\alpha=0$, $\theta(0)=0$, $\phi(0)\rightarrow\phi$, $\theta(T/2)=\pi/2$, and $\phi(T/2)\rightarrow\phi$, the system population can be first transferred from $|0\rangle$ to $|1\rangle$. Next, it can be transferred as $|1\rangle\rightarrow|e\rangle\rightarrow|0\rangle$ via the path $|\mu_2(t)\rangle$ with $\beta=0$, $\theta(T/2)=\phi(T/2)=0$, $\theta(T)\rightarrow\theta$, $\phi(T)=\pi/2$, $\theta(3T/2)=\pi/2$, and $\phi(3T/2)=0$.

For the three-level system, the cyclic loop of population transfer in either direction requires a full-rank nonadiabatic evolution operator. With our universal protocol, the loop is based on the manipulation over $3$ degrees of freedom of the system, i.e., the Rabi frequencies of the three driving fields. In contrast, the counterdiabatic driving method~\cite{Chen2010Shortcut} has to employ the Rabi frequencies of five driving fields to achieve a full-rank evolution. Compared to our configuration in Fig.~\ref{model}, two extra driving fields are provided in the transitions $|0\rangle\leftrightarrow|e\rangle$ and $|1\rangle\leftrightarrow|e\rangle$. Then our protocol offers a resource-saving technique for full-rank nonadiabatic evolution.

\begin{figure}[htbp]
\centering
\includegraphics[width=0.9\linewidth]{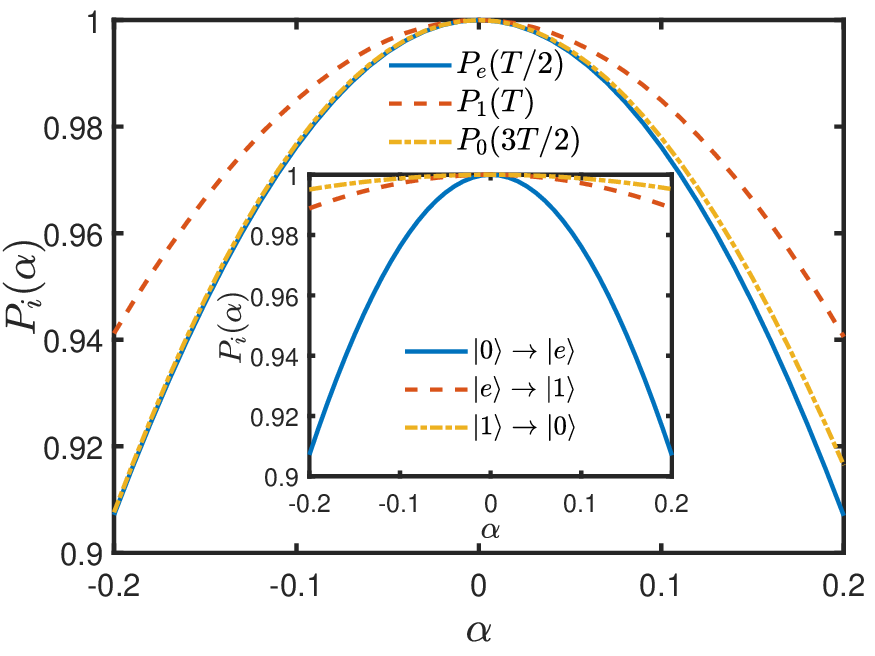}
\caption{State populations $P_i(\alpha)$ as a function of the local dimensionless error $\alpha$ in the Rabi-frequency $\Omega_0(t)$ for the driving between $|0\rangle$ and $|e\rangle$. The other parameters are set the same as in Fig.~\ref{transfer}.}\label{error}
\end{figure}

To test the robustness of the cyclic path against the parametric fluctuation, we consider that the driving intensity $\Omega_0(t)$ between $|0\rangle$ and $|e\rangle$ deviates from Eq.~(\ref{condition}), i.e., $\Omega_0(t)\rightarrow(1+\alpha)\Omega_0(t)$. The local dimensionless error $\alpha$ influences both stages represented by Eqs.~(\ref{Stage1}) and (\ref{Stage2}) in the cyclic loops. The performance of the practical evolution can be evaluated by the overlap between the system state and the target states. It has almost the same definition as the state population $P_i(\alpha)\equiv\langle i|\psi_{\alpha}(t)\rangle\langle\psi_{\alpha}(t)|i\rangle$, where $t$ is fixed as $T/2$, $T$, and $3T/2$ for different target states indicated by $i=0, 1, e$, respectively. $|\psi_{\alpha}(t)\rangle$ is obtained using the Schr\"odinger equation with the error Hamiltonian. In Fig.~\ref{error}, we demonstrate $P_i(\alpha)$ as a function of $\alpha$ at the desired moments for the population transfer. In the main frame and in the subframe, the error presents in the whole process and only in the specified process, respectively. The target-state population is found to be more sensitive to $\alpha$ during the process $|0\rangle\rightarrow|e\rangle$ than the other two. $P_i(\alpha)$ generally decreases with the error magnitude. However, it is still above $0.91$, even if $|\alpha|=0.2$.

\subsection{Four-level system}\label{fourlevelsystem}

\begin{figure}[htbp]
\centering
\includegraphics[width=0.7\linewidth]{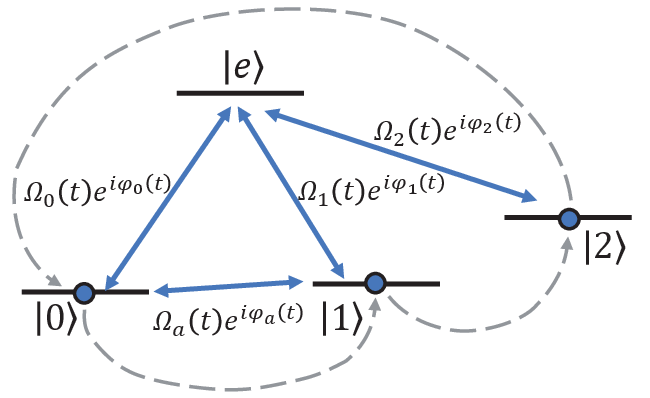}
\caption{Sketch of a four-level system, in which the transitions $|e\rangle\leftrightarrow|i\rangle$, $i=0,1,2$, and $|0\rangle\leftrightarrow|1\rangle$ are driven by the resonant driving fields. The cyclic population transfer $|2\rangle\rightarrow|0\rangle\rightarrow|1\rangle\rightarrow|2\rangle$ is marked by the gray dashed line.}\label{modelfour}
\end{figure}

In Fig.~\ref{modelfour}, we consider a four-level system under four resonant driving fields with Rabi frequencies $\Omega_i(t)$ and phases $\varphi_i(t)$, $i=0,1,2,a$. Although this configuration is not the most general one, it is sufficient to realize the cyclic transfer among three levels: $|0\rangle$, $|1\rangle$, and $|2\rangle$. The Hamiltonian can be written as
\begin{equation}\label{Hamfour}
\begin{aligned}
&H(t)=\Omega_0(t)e^{i\varphi_0(t)}|0\rangle\langle e|+\Omega_1(t)e^{i\varphi_1(t)}|1\rangle\langle e|\\
&+\Omega_2(t)e^{i\varphi_2(t)}|2\rangle\langle e|+\Omega_a(t)e^{i\varphi_a(t)}|0\rangle\langle 1|+{\rm H.c.}.
\end{aligned}
\end{equation}
By extending the three-dimensional case in Eq.~(\ref{auxbasis}) or its specific form in Eq.~(\ref{auxbasisLewis}), a four-dimensional ancillary picture can be constructed as
\begin{equation}\label{auxFour}
\begin{aligned}
|\mu_0(t)\rangle&=\cos\theta(t)|0\rangle-\sin\theta(t)|1\rangle,\\
|\mu_1(t)\rangle&=\sin\phi(t)\sin\theta(t)|0\rangle+\cos\phi(t)|e\rangle\\
&+\sin\phi(t)\cos\theta(t)|1\rangle,\\
|\mu_2(t)\rangle&=\cos\chi(t)\cos\phi(t)\sin\theta(t)|0\rangle-\cos\chi(t)\sin\phi(t)|e\rangle\\
&+\cos\chi(t)\cos\phi(t)\cos\theta(t)|1\rangle+\sin\chi(t)|2\rangle,\\
|\mu_3(t)\rangle&=\sin\chi(t)\cos\phi(t)\sin\theta(t)|0\rangle-\sin\chi(t)\sin\phi(t)|e\rangle\\
&+\sin\chi(t)\cos\phi(t)\cos\theta(t)|1\rangle-\cos\chi(t)|2\rangle,
\end{aligned}
\end{equation}
where $\chi(t)$, $\phi(t)$, and $\theta(t)$ are time-dependent parameters. Due to the lack of a sufficient number of degrees of freedom, i.e., the absence of the driving fields on the transitions $|0\rangle\leftrightarrow|2\rangle$ and $|1\rangle\leftrightarrow|2\rangle$, only two of the four ancillary states, e.g., $\{|\mu_0(t)\rangle, |\mu_2(t)\rangle\}$ or $\{|\mu_0(t)\rangle, |\mu_3(t)\rangle\}$, can satisfy the von Neumann equation~(\ref{von}) under the same parametric setting. Thus, if $|\mu_0(t)\rangle$ and $|\mu_2(t)\rangle$ are chosen to span a two-dimensional subspace for the controllable dynamics, then one can construct the cyclic population transfer along the repeated loop $|2\rangle\rightarrow|0\rangle\rightarrow|1\rangle\rightarrow|2\rangle$. In particular, the von Neumann equation~(\ref{von}) for $|\mu_0(t)\rangle$ and $|\mu_2(t)\rangle$ gives rise to constant phases as
\begin{equation}\label{phasefour}
\varphi_0=\varphi_1=\varphi_a=\frac{\pi}{2}, \quad \varphi_2=-\frac{\pi}{2}
\end{equation}
and the Rabi-frequencies as
\begin{equation}\label{RabiFour}
\begin{aligned}
&\Omega_0(t)=\Omega(t)\sin\theta(t),\\
&\Omega_1(t)=\Omega(t)\cos\theta(t),\\
&\Omega_2(t)=\frac{\dot{\chi}(t)}{\sin\phi(t)},\\
&\Omega_a(t)=\dot{\theta}(t),
\end{aligned}
\end{equation}
where $\Omega(t)=\dot{\chi}(t)\tan\chi(t)\cot\phi(t)+\dot{\phi}(t)$. These conditions ensure that there is no transition between $|\mu_0(t)\rangle$ and $|\mu_2(t)\rangle$ during the whole evolution. Moreover, it is found that $|e\rangle$ can not be fully populated due to a nonvanishing $\Omega(t)$ and the boundary conditions of $|\mu_0(t)\rangle$ and $|\mu_2(t)\rangle$ for cyclic transfer.

A completed loop $|2\rangle\rightarrow|0\rangle\rightarrow|1\rangle\rightarrow|2\rangle$ can be divided into three stages. In stage $1$, the system follows the path $|\mu_2(t)\rangle$ in Eq.~(\ref{auxFour}), and the parameters can be set as
\begin{equation}\label{stagefour1}
\begin{aligned}
\phi(t)&=\frac{\pi}{2}\cos\frac{\pi t}{T},\\
\chi(t)&=\frac{\pi}{2}\left[\cos\phi(t)-1\right],\\
\theta(t)&=-\frac{\pi}{2}\sin\frac{\pi t}{T}.
\end{aligned}
\end{equation}
Then the population is transferred from $|2\rangle$ to $|0\rangle$ at $t=T/2$. In stage $2$, one can employ the path $|\mu_0(t)\rangle$ in Eq.~(\ref{auxFour}), and the parameters are set as
\begin{equation}\label{stagefour2}
\begin{aligned}
\phi(t)&=\frac{\pi}{2}\cos\frac{\pi t}{T},\\
\chi(t)&=\frac{\pi}{2}\left[\cos\phi(t)-1\right],\\
\theta(t)&=\chi(t),
\end{aligned}
\end{equation}
by which the population on $|0\rangle$ is transferred to $|1\rangle$ at $t=T$. In stage $3$, the system population can be moved back to $|2\rangle$ through the path $|\mu_2(t)\rangle$ with
\begin{equation}\label{stagefour3}
\begin{aligned}
\phi(t)&=\frac{\pi}{2}\sin\frac{\pi t}{T},\\
\chi(t)&=\frac{\pi}{2}\left[\cos\phi(t)-1\right],\\
\theta(t)&=\chi(t).
\end{aligned}
\end{equation}
When $t=3T/2$, one loop of cyclic population transfer is completed.

\begin{table*}[htbp]
\centering
\caption{Comparison of our protocol and existing ones with respect to state transfer in a $\Lambda$-type three-level system. NHQT, nonadiabatic holonomic quantum transformation, e.g., nonadiabatic holonomic quantum computation combined with spin-echo or pulse-shaping methods~\cite{Liu2019Plug}; LRT, Lewis-Riesenfeld theory for invariants; CD, counterdiabatic driving; CDD, counterdiabatic driving in dressed states.}\label{table}
\begin{threeparttable}
\begin{tabular}{ccccccc}
\hline \hline
\multicolumn{1}{c}{}&\multicolumn{6}{c}{Protocol}\\
\hline
\multicolumn{1}{c}{}&\multicolumn{1}{c}{Our protocol}&\multicolumn{1}{c}{STIRAP}&\multicolumn{1}{c}{NHQT}&
\multicolumn{1}{c}{LRT}&
\multicolumn{1}{c}{CD}&
\multicolumn{1}{c}{CDD}\\
\hline
Nonadiabatic& $\checkmark$ &$\times$ & \checkmark & \checkmark & \checkmark&\checkmark\\
State transfer via $|\mu_1(t)\rangle$ or $|\mu_2(t)\rangle$ in Eq.~(\ref{auxbasis})& \checkmark & $\times$ & \checkmark & \checkmark &$\times$&\checkmark\\
State transfer via $|\mu_0(t)\rangle$ in Eq.~(\ref{auxbasis})& \checkmark & \checkmark & $\times$ & $\times$ & \checkmark& $\times$\\
Cyclic transfer & \checkmark & $\times$ & $\times$ & $\times$ & $\times$&$\times$\\
Full rank & \checkmark & $\checkmark$ & $\times$ & $\times$ & $\times$&$\times$\\
References & This work & \cite{Vitanov2017Stiumlated,Pillet1993Adiabatic,Phillips1998Nobel,Garc2003Quantum,Daems2007Adiabatic}& \cite{Sjoqvist2012Nonadiabatic,Liu2019Plug} &~\cite{Chen2011Lewis,Chen2012Engineering,Guery2019Shortcuts} & \cite{Chen2010Shortcut,Antti2019Superadiabatic,Guery2019Shortcuts}&\cite{Baksic2016Speeding,Li2016Shortcut,Guery2019Shortcuts}\\
\hline \hline
\end{tabular}
\end{threeparttable}
\end{table*}

\begin{figure}[htbp]
\centering
\includegraphics[width=0.9\linewidth]{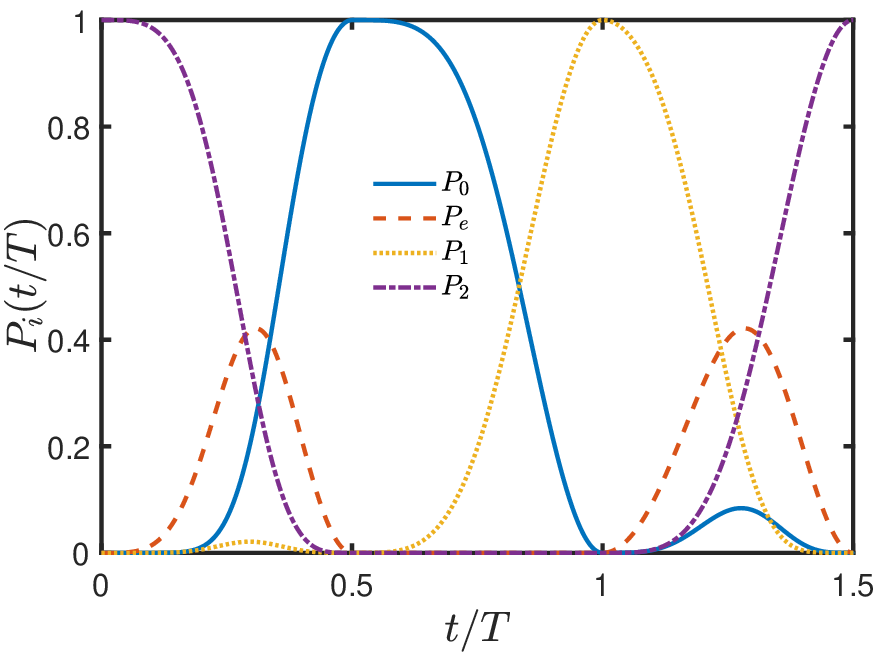}
\caption{State populations $P_i(t)$, $i=1,2,3,a$, versus the evolution time $t/T$ for the cyclic population transfer. The Rabi-frequencies $\Omega_i(t)$ and the phases $\varphi_i(t)$, $i=1,2,3,a$, are set according to Eqs.~(\ref{phasefour}) and (\ref{RabiFour}) with the parameters $\chi(t)$, $\phi(t)$, and $\theta(t)$ given in Eqs.~(\ref{stagefour1}), (\ref{stagefour2}), and (\ref{stagefour3}).}\label{transferfour}
\end{figure}

The population dynamics of the four levels $P_i(t)$ are demonstrated in Fig.~\ref{transferfour}, along the paths of the von Neumann equation~(\ref{von}). It is found that the initial population on the state $|2\rangle$ can be completely transferred to $|0\rangle$ when $t=T/2$. During this transfer process, the states $|1\rangle$ and $|e\rangle$ are temporally populated, i.e., $P_1(t)=0.02$ and $P_e(t)=0.42$ when $t=0.31T$, due to the fact that $|\mu_2(t)\rangle$ is a superposed state of all four levels. Then at $t=T$, the population on the state $|0\rangle$ can be completely transferred to $|1\rangle$ via the path $|\mu_0(t)\rangle$. It is interesting to find that stage $2$ is the same as that for the three-level system. From $t=T$ to $t=3T/2$, the population on the state $|1\rangle$ is transferred back to $|2\rangle$ again along the path $|\mu_2(t)\rangle$.

When the system is initially prepared in $|0\rangle$ or $|1\rangle$, the preceding cyclic transfer can be performed with only two stages. For example, if the cyclic loop is $|0\rangle\rightarrow|1\rangle\rightarrow|2\rangle\rightarrow|0\rangle$, then the first and second stages employ the paths $|\mu_0(t)\rangle$ and $|\mu_2(t)\rangle$, respectively. Along the path $|\mu_0(t)\rangle$ with $\theta(0)=0$ and $\theta(T/2)=\pi/2$, the system population can be transferred from $|0\rangle$ to $|1\rangle$. Then the population transfer $|1\rangle\rightarrow|2\rangle\rightarrow|0\rangle$ can be achieved along the path $|\mu_2(t)\rangle$ with $\chi(T/2)=0$, $\phi(T/2)=0$, $\theta(T/2)=0$, $\chi(T)=\pi/2$, $\phi(T)\rightarrow\phi$, $\theta(T)\rightarrow\theta$, $\chi(3T/2)=0$, $\phi(3T/2)=0$, and $\theta(3T/2)=\pi/2$.

\section{Discussion}\label{discussion}

We summarize in Table~\ref{table} the common and distinct points between our universal protocol and the existing ones with respect to the state transfer in a $\Lambda$ system. It is a popular and pedagogical example for unveiling the control capability with only the three driving fields, as in Fig.~\ref{model}. In addition to the adiabatic evolution path constructed in STIRAP, one can order a stable and fast evolution path using our nonadiabatic control method. Our universal framework covers NHQT, LRT, CD, and CDD methods. The full-rank evolution operator obtained with our universal protocol allows us to choose an arbitrary ancillary state as the state-transfer path. This versatility is superior to CD on $|\mu_0(t)\rangle$ and NHQT, LRT, and CDD on $|\mu_1(t)\rangle$ or $|\mu_2(t)\rangle$. One can also appreciate our protocol through its application in cyclic population transfer.

For a finite-dimensional quantum system, our protocol is one shot with respect to the rotation to the ancillary picture. In contrast, double unitary rotations have to be performed on the system Hamiltonian under the modified CD or CDD method~\cite{Baksic2016Speeding,Li2016Shortcut,Yue2018Fast}. In the first rotated picture, the Hamiltonian can be diagonal with the assistance of a possibly unphysical driving field. Then another rotation is designed to cancel the prohibited pumping between particular levels.

\emph{A brief recipe for our protocol.} Through the von Neumann equation~(\ref{von}), one can be equipped with at least one stable and fast evolution path for a quantum system, provided that the time-dependent Hamiltonian has a sufficient number of degrees of freedom or tunable parameters to avoid ending up with dark modes. The boundary conditions of the tunable parameters of the nonadiabatic evolution path and the rank of the time-evolution operator~(\ref{U}) determine which state can be achieved.

\section{Conclusion}\label{conclusion}

In conclusion, we presented a forward-engineering framework for quantum control over time-dependent systems within an ancillary picture. The instantaneous ancillary basis states that satisfy the von Neumann equation under the system Hamiltonian provide nonadiabatic evolution paths for the system of interest. No transition occurs among the eligible ancillary states since the system Hamiltonian is diagonal in them. One can achieve a full-rank evolution operator under a time-dependent Hamiltonian with a sufficient number of degrees of freedom. This means that the Hamiltonian is fully diagonal in the parametric ancillary picture and the time-dependent Schr\"odinger equation can be exactly solved. Our universal prescription can unify many popular state-transfer protocols, including nonadiabatic holonomic transformation, the Lewis-Riesenfeld theory for invariants, and counterdiabatic driving methods. With a clear motivation and a convenient derivation, our protocol allows us to construct multiple stable and fast paths linking two arbitrary states of time-dependent quantum systems. We actually proved that STA and holonomic transformation are the same thing.

More than highlighting the common basic elements of the existing control protocols, our framework works for certain tasks beyond their reach or greatly reduces the complexity and performance cost. Protocols for cyclic population transfer in a three-level system and a four-level system are designed in our framework, demonstrating its advantage in exploiting the potential of nonadiabatic control.

\section*{Acknowledgments}

We acknowledge grant support from the National Natural Science Foundation of China (Grant No. 11974311).

\bibliographystyle{apsrevlong}
\bibliography{ref}

\begin{thebibliography}{41}%
\makeatletter
\providecommand \@ifxundefined [1]{%
 \@ifx{#1\undefined}
}%
\providecommand \@ifnum [1]{%
 \ifnum #1\expandafter \@firstoftwo
 \else \expandafter \@secondoftwo
 \fi
}%
\providecommand \@ifx [1]{%
 \ifx #1\expandafter \@firstoftwo
 \else \expandafter \@secondoftwo
 \fi
}%
\providecommand \natexlab [1]{#1}%
\providecommand \enquote  [1]{``#1''}%
\providecommand \bibnamefont  [1]{#1}%
\providecommand \bibfnamefont [1]{#1}%
\providecommand \citenamefont [1]{#1}%
\providecommand \href@noop [0]{\@secondoftwo}%
\providecommand \href [0]{\begingroup \@sanitize@url \@href}%
\providecommand \@href[1]{\@@startlink{#1}\@@href}%
\providecommand \@@href[1]{\endgroup#1\@@endlink}%
\providecommand \@sanitize@url [0]{\catcode `\\12\catcode `\$12\catcode
  `\&12\catcode `\#12\catcode `\^12\catcode `\_12\catcode `\%12\relax}%
\providecommand \@@startlink[1]{}%
\providecommand \@@endlink[0]{}%
\providecommand \url  [0]{\begingroup\@sanitize@url \@url }%
\providecommand \@url [1]{\endgroup\@href {#1}{\urlprefix }}%
\providecommand \urlprefix  [0]{URL }%
\providecommand \Eprint [0]{\href }%
\providecommand \doibase [0]{http://dx.doi.org/}%
\providecommand \selectlanguage [0]{\@gobble}%
\providecommand \bibinfo  [0]{\@secondoftwo}%
\providecommand \bibfield  [0]{\@secondoftwo}%
\providecommand \translation [1]{[#1]}%
\providecommand \BibitemOpen [0]{}%
\providecommand \bibitemStop [0]{}%
\providecommand \bibitemNoStop [0]{.\EOS\space}%
\providecommand \EOS [0]{\spacefactor3000\relax}%
\providecommand \BibitemShut  [1]{\csname bibitem#1\endcsname}%
\let\auto@bib@innerbib\@empty
\bibitem [{\citenamefont {Kr\'al}\ \emph {et~al.}(2007)\citenamefont {Kr\'al},
  \citenamefont {Thanopulos},\ and\ \citenamefont
  {Shapiro}}]{Kral2007Colloquium}%
  \BibitemOpen
  \bibfield  {author} {\bibinfo {author} {\bibfnamefont {P.}~\bibnamefont
  {Kr\'al}}, \bibinfo {author} {\bibfnamefont {I.}~\bibnamefont {Thanopulos}},
  \ and\ \bibinfo {author} {\bibfnamefont {M.}~\bibnamefont {Shapiro}},\
  }\bibfield  {title} {\emph {\bibinfo {title} {Colloquium: Coherently
  controlled adiabatic passage},\ }}\href {\doibase 10.1103/RevModPhys.79.53}
  {\bibfield  {journal} {\bibinfo  {journal} {Rev. Mod. Phys.}\ }\textbf
  {\bibinfo {volume} {79}},\ \bibinfo {pages} {53} (\bibinfo {year}
  {2007})}\BibitemShut {NoStop}%
\bibitem [{\citenamefont {Zanardi}\ and\ \citenamefont
  {Rasetti}(1999)}]{Zanardi1999Holonomic}%
  \BibitemOpen
  \bibfield  {author} {\bibinfo {author} {\bibfnamefont {P.}~\bibnamefont
  {Zanardi}}\ and\ \bibinfo {author} {\bibfnamefont {M.}~\bibnamefont
  {Rasetti}},\ }\bibfield  {title} {\emph {\bibinfo {title} {Holonomic quantum
  computation},\ }}\href {\doibase 10.1016/S0375-9601(99)00803-8} {\bibfield
  {journal} {\bibinfo  {journal} {Phys. Lett. A}\ }\textbf {\bibinfo {volume}
  {264}},\ \bibinfo {pages} {94} (\bibinfo {year} {1999})}\BibitemShut
  {NoStop}%
\bibitem [{\citenamefont {Bergmann}\ \emph {et~al.}(1998)\citenamefont
  {Bergmann}, \citenamefont {Theuer},\ and\ \citenamefont
  {Shore}}]{Bergmann1998Coherent}%
  \BibitemOpen
  \bibfield  {author} {\bibinfo {author} {\bibfnamefont {K.}~\bibnamefont
  {Bergmann}}, \bibinfo {author} {\bibfnamefont {H.}~\bibnamefont {Theuer}}, \
  and\ \bibinfo {author} {\bibfnamefont {B.~W.}\ \bibnamefont {Shore}},\
  }\bibfield  {title} {\emph {\bibinfo {title} {Coherent population transfer
  among quantum states of atoms and molecules},\ }}\href {\doibase
  10.1103/RevModPhys.70.1003} {\bibfield  {journal} {\bibinfo  {journal} {Rev.
  Mod. Phys.}\ }\textbf {\bibinfo {volume} {70}},\ \bibinfo {pages} {1003}
  (\bibinfo {year} {1998})}\BibitemShut {NoStop}%
\bibitem [{\citenamefont {Gisin}\ and\ \citenamefont
  {Thew}(2007)}]{Gisin2007Quantum}%
  \BibitemOpen
  \bibfield  {author} {\bibinfo {author} {\bibfnamefont {N.}~\bibnamefont
  {Gisin}}\ and\ \bibinfo {author} {\bibfnamefont {R.}~\bibnamefont {Thew}},\
  }\bibfield  {title} {\emph {\bibinfo {title} {Quantum communication},\
  }}\href {\doibase 10.1038/nphoton.2007.22} {\bibfield  {journal} {\bibinfo
  {journal} {Nat. Photonics}\ }\textbf {\bibinfo {volume} {1}},\ \bibinfo
  {pages} {165} (\bibinfo {year} {2007})}\BibitemShut {NoStop}%
\bibitem [{\citenamefont {Kimble}(2008)}]{Kimble2008Quantum}%
  \BibitemOpen
  \bibfield  {author} {\bibinfo {author} {\bibfnamefont {H.~J.}\ \bibnamefont
  {Kimble}},\ }\bibfield  {title} {\emph {\bibinfo {title} {The quantum
  internet},\ }}\href {\doibase 10.1038/nature07127} {\bibfield  {journal}
  {\bibinfo  {journal} {Nature}\ }\textbf {\bibinfo {volume} {453}},\ \bibinfo
  {pages} {1023} (\bibinfo {year} {2008})}\BibitemShut {NoStop}%
\bibitem [{\citenamefont {Dong}\ \emph {et~al.}(2021)\citenamefont {Dong},
  \citenamefont {Zhuang}, \citenamefont {Economou},\ and\ \citenamefont
  {Barnes}}]{Dong2021Doubly}%
  \BibitemOpen
  \bibfield  {author} {\bibinfo {author} {\bibfnamefont {W.}~\bibnamefont
  {Dong}}, \bibinfo {author} {\bibfnamefont {F.}~\bibnamefont {Zhuang}},
  \bibinfo {author} {\bibfnamefont {S.~E.}\ \bibnamefont {Economou}}, \ and\
  \bibinfo {author} {\bibfnamefont {E.}~\bibnamefont {Barnes}},\ }\bibfield
  {title} {\emph {\bibinfo {title} {Doubly geometric quantum control},\ }}\href
  {\doibase 10.1103/PRXQuantum.2.030333} {\bibfield  {journal} {\bibinfo
  {journal} {PRX Quantum}\ }\textbf {\bibinfo {volume} {2}},\ \bibinfo {pages}
  {030333} (\bibinfo {year} {2021})}\BibitemShut {NoStop}%
\bibitem [{\citenamefont {Berry}(1984)}]{Berry1984Quantal}%
  \BibitemOpen
  \bibfield  {author} {\bibinfo {author} {\bibfnamefont {M.~V.}\ \bibnamefont
  {Berry}},\ }\bibfield  {title} {\emph {\bibinfo {title} {Quantal phase
  factors accompanying adiabatic changes},\ }}\href {\doibase
  10.1007/BF03046050} {\bibfield  {journal} {\bibinfo  {journal} {Proc. R. Soc.
  A}\ }\textbf {\bibinfo {volume} {392}},\ \bibinfo {pages} {45} (\bibinfo
  {year} {1984})}\BibitemShut {NoStop}%
\bibitem [{\citenamefont {Aharonov}\ and\ \citenamefont
  {Anandan}(1987)}]{Anharonov1987Phase}%
  \BibitemOpen
  \bibfield  {author} {\bibinfo {author} {\bibfnamefont {Y.}~\bibnamefont
  {Aharonov}}\ and\ \bibinfo {author} {\bibfnamefont {J.}~\bibnamefont
  {Anandan}},\ }\bibfield  {title} {\emph {\bibinfo {title} {Phase change
  during a cyclic quantum evolution},\ }}\href {\doibase
  10.1103/PhysRevLett.58.1593} {\bibfield  {journal} {\bibinfo  {journal}
  {Phys. Rev. Lett.}\ }\textbf {\bibinfo {volume} {58}},\ \bibinfo {pages}
  {1593} (\bibinfo {year} {1987})}\BibitemShut {NoStop}%
\bibitem [{\citenamefont {Wilczek}\ and\ \citenamefont
  {Zee}(1984)}]{Wilczek1984Appearance}%
  \BibitemOpen
  \bibfield  {author} {\bibinfo {author} {\bibfnamefont {F.}~\bibnamefont
  {Wilczek}}\ and\ \bibinfo {author} {\bibfnamefont {A.}~\bibnamefont {Zee}},\
  }\bibfield  {title} {\emph {\bibinfo {title} {Appearance of gauge structure
  in simple dynamical systems},\ }}\href {\doibase 10.1103/PhysRevLett.52.2111}
  {\bibfield  {journal} {\bibinfo  {journal} {Phys. Rev. Lett.}\ }\textbf
  {\bibinfo {volume} {52}},\ \bibinfo {pages} {2111} (\bibinfo {year}
  {1984})}\BibitemShut {NoStop}%
\bibitem [{\citenamefont {Anandan}(1988)}]{Anandan1988Non}%
  \BibitemOpen
  \bibfield  {author} {\bibinfo {author} {\bibfnamefont {J.}~\bibnamefont
  {Anandan}},\ }\bibfield  {title} {\emph {\bibinfo {title} {Non-adiabatic
  non-abelian geometric phase},\ }}\href {\doibase
  https://doi.org/10.1016/0375-9601(88)91010-9} {\bibfield  {journal} {\bibinfo
   {journal} {Phys. Lett. A}\ }\textbf {\bibinfo {volume} {133}},\ \bibinfo
  {pages} {171} (\bibinfo {year} {1988})}\BibitemShut {NoStop}%
\bibitem [{\citenamefont {De~Chiara}\ and\ \citenamefont
  {Palma}(2003)}]{De2003Berry}%
  \BibitemOpen
  \bibfield  {author} {\bibinfo {author} {\bibfnamefont {G.}~\bibnamefont
  {De~Chiara}}\ and\ \bibinfo {author} {\bibfnamefont {G.~M.}\ \bibnamefont
  {Palma}},\ }\bibfield  {title} {\emph {\bibinfo {title} {Berry phase for a
  spin $1/2$ particle in a classical fluctuating field},\ }}\href {\doibase
  10.1103/PhysRevLett.91.090404} {\bibfield  {journal} {\bibinfo  {journal}
  {Phys. Rev. Lett.}\ }\textbf {\bibinfo {volume} {91}},\ \bibinfo {pages}
  {090404} (\bibinfo {year} {2003})}\BibitemShut {NoStop}%
\bibitem [{\citenamefont {Leek}\ \emph {et~al.}(2007)\citenamefont {Leek},
  \citenamefont {Fink}, \citenamefont {Blais}, \citenamefont {Bianchetti},
  \citenamefont {G\"oppl}, \citenamefont {Gambetta}, \citenamefont {Schuster},
  \citenamefont {Frunzio}, \citenamefont {Schoelkopf},\ and\ \citenamefont
  {Wallraff}}]{Leek2007Observation}%
  \BibitemOpen
  \bibfield  {author} {\bibinfo {author} {\bibfnamefont {P.~J.}\ \bibnamefont
  {Leek}}, \bibinfo {author} {\bibfnamefont {J.~M.}\ \bibnamefont {Fink}},
  \bibinfo {author} {\bibfnamefont {A.}~\bibnamefont {Blais}}, \bibinfo
  {author} {\bibfnamefont {R.}~\bibnamefont {Bianchetti}}, \bibinfo {author}
  {\bibfnamefont {M.}~\bibnamefont {G\"oppl}}, \bibinfo {author} {\bibfnamefont
  {J.~M.}\ \bibnamefont {Gambetta}}, \bibinfo {author} {\bibfnamefont {D.~I.}\
  \bibnamefont {Schuster}}, \bibinfo {author} {\bibfnamefont {L.}~\bibnamefont
  {Frunzio}}, \bibinfo {author} {\bibfnamefont {R.~J.}\ \bibnamefont
  {Schoelkopf}}, \ and\ \bibinfo {author} {\bibfnamefont {A.}~\bibnamefont
  {Wallraff}},\ }\bibfield  {title} {\emph {\bibinfo {title} {Observation of
  berry’s phase in a solid-state qubit},\ }}\href {\doibase
  10.1126/science.1149858} {\bibfield  {journal} {\bibinfo  {journal}
  {Science}\ }\textbf {\bibinfo {volume} {318}},\ \bibinfo {pages} {1889}
  (\bibinfo {year} {2007})}\BibitemShut {NoStop}%
\bibitem [{\citenamefont {Filipp}\ \emph {et~al.}(2009)\citenamefont {Filipp},
  \citenamefont {Klepp}, \citenamefont {Hasegawa}, \citenamefont
  {Plonka-Spehr}, \citenamefont {Schmidt}, \citenamefont {Geltenbort},\ and\
  \citenamefont {Rauch}}]{Filipp2009Experimental}%
  \BibitemOpen
  \bibfield  {author} {\bibinfo {author} {\bibfnamefont {S.}~\bibnamefont
  {Filipp}}, \bibinfo {author} {\bibfnamefont {J.}~\bibnamefont {Klepp}},
  \bibinfo {author} {\bibfnamefont {Y.}~\bibnamefont {Hasegawa}}, \bibinfo
  {author} {\bibfnamefont {C.}~\bibnamefont {Plonka-Spehr}}, \bibinfo {author}
  {\bibfnamefont {U.}~\bibnamefont {Schmidt}}, \bibinfo {author} {\bibfnamefont
  {P.}~\bibnamefont {Geltenbort}}, \ and\ \bibinfo {author} {\bibfnamefont
  {H.}~\bibnamefont {Rauch}},\ }\bibfield  {title} {\emph {\bibinfo {title}
  {Experimental demonstration of the stability of berry's phase for a
  spin-$1/2$ particle},\ }}\href {\doibase 10.1103/PhysRevLett.102.030404}
  {\bibfield  {journal} {\bibinfo  {journal} {Phys. Rev. Lett.}\ }\textbf
  {\bibinfo {volume} {102}},\ \bibinfo {pages} {030404} (\bibinfo {year}
  {2009})}\BibitemShut {NoStop}%
\bibitem [{\citenamefont {Cohen}\ \emph {et~al.}(2019)\citenamefont {Cohen},
  \citenamefont {Larocque}, \citenamefont {Bouchard}, \citenamefont
  {Nejadsattari}, \citenamefont {Gefen},\ and\ \citenamefont
  {Karimi}}]{Cohen2019Geometric}%
  \BibitemOpen
  \bibfield  {author} {\bibinfo {author} {\bibfnamefont {E.}~\bibnamefont
  {Cohen}}, \bibinfo {author} {\bibfnamefont {H.}~\bibnamefont {Larocque}},
  \bibinfo {author} {\bibfnamefont {F.}~\bibnamefont {Bouchard}}, \bibinfo
  {author} {\bibfnamefont {F.}~\bibnamefont {Nejadsattari}}, \bibinfo {author}
  {\bibfnamefont {Y.}~\bibnamefont {Gefen}}, \ and\ \bibinfo {author}
  {\bibfnamefont {E.}~\bibnamefont {Karimi}},\ }\bibfield  {title} {\emph
  {\bibinfo {title} {Geometric phase from {Aharonov–Bohm to
  Pancharatnam–Berry} and beyond},\ }}\href {\doibase
  10.1038/s42254-019-0071-1} {\bibfield  {journal} {\bibinfo  {journal} {Nat.
  Rev. Phys.}\ }\textbf {\bibinfo {volume} {1}},\ \bibinfo {pages} {437}
  (\bibinfo {year} {2019})}\BibitemShut {NoStop}%
\bibitem [{\citenamefont {Born}\ and\ \citenamefont
  {Fock}(1928)}]{Born1928Beweis}%
  \BibitemOpen
  \bibfield  {author} {\bibinfo {author} {\bibfnamefont {M.}~\bibnamefont
  {Born}}\ and\ \bibinfo {author} {\bibfnamefont {V.}~\bibnamefont {Fock}},\
  }\bibfield  {title} {\emph {\bibinfo {title} {Beweis des adiabatensatzes},\
  }}\href {\doibase 10.1007/BF01343193} {\bibfield  {journal} {\bibinfo
  {journal} {Z. Phys.}\ }\textbf {\bibinfo {volume} {51}},\ \bibinfo {pages}
  {165} (\bibinfo {year} {1928})}\BibitemShut {NoStop}%
\bibitem [{\citenamefont {Jones}\ \emph {et~al.}(2000)\citenamefont {Jones},
  \citenamefont {Vedral}, \citenamefont {Ekert},\ and\ \citenamefont
  {Castagnoli}}]{Jones2000Geometric}%
  \BibitemOpen
  \bibfield  {author} {\bibinfo {author} {\bibfnamefont {J.~A.}\ \bibnamefont
  {Jones}}, \bibinfo {author} {\bibfnamefont {V.}~\bibnamefont {Vedral}},
  \bibinfo {author} {\bibfnamefont {A.}~\bibnamefont {Ekert}}, \ and\ \bibinfo
  {author} {\bibfnamefont {G.}~\bibnamefont {Castagnoli}},\ }\bibfield  {title}
  {\emph {\bibinfo {title} {Geometric quantum computation using nuclear
  magnetic resonance},\ }}\href {\doibase 10.1038/35002528} {\bibfield
  {journal} {\bibinfo  {journal} {Nature}\ }\textbf {\bibinfo {volume} {403}},\
  \bibinfo {pages} {869} (\bibinfo {year} {2000})}\BibitemShut {NoStop}%
\bibitem [{\citenamefont {Duan}\ \emph {et~al.}(2001)\citenamefont {Duan},
  \citenamefont {Cirac},\ and\ \citenamefont {Zoller}}]{Duan2001Geometric}%
  \BibitemOpen
  \bibfield  {author} {\bibinfo {author} {\bibfnamefont {L.-M.}\ \bibnamefont
  {Duan}}, \bibinfo {author} {\bibfnamefont {J.~I.}\ \bibnamefont {Cirac}}, \
  and\ \bibinfo {author} {\bibfnamefont {P.}~\bibnamefont {Zoller}},\
  }\bibfield  {title} {\emph {\bibinfo {title} {Geometric manipulation of
  trapped ions for quantum computation},\ }}\href {\doibase
  10.1126/science.1058835} {\bibfield  {journal} {\bibinfo  {journal}
  {Science}\ }\textbf {\bibinfo {volume} {292}},\ \bibinfo {pages} {1695}
  (\bibinfo {year} {2001})}\BibitemShut {NoStop}%
\bibitem [{\citenamefont {Sjöqvist}\ \emph {et~al.}(2012)\citenamefont
  {Sjöqvist}, \citenamefont {Tong}, \citenamefont {Andersson}, \citenamefont
  {Hessmo}, \citenamefont {Johansson},\ and\ \citenamefont
  {Singh}}]{Sjoqvist2012Nonadiabatic}%
  \BibitemOpen
  \bibfield  {author} {\bibinfo {author} {\bibfnamefont {E.}~\bibnamefont
  {Sjöqvist}}, \bibinfo {author} {\bibfnamefont {D.~M.}\ \bibnamefont {Tong}},
  \bibinfo {author} {\bibfnamefont {L.~M.}\ \bibnamefont {Andersson}}, \bibinfo
  {author} {\bibfnamefont {B.}~\bibnamefont {Hessmo}}, \bibinfo {author}
  {\bibfnamefont {M.}~\bibnamefont {Johansson}}, \ and\ \bibinfo {author}
  {\bibfnamefont {K.}~\bibnamefont {Singh}},\ }\bibfield  {title} {\emph
  {\bibinfo {title} {Non-adiabatic holonomic quantum computation},\ }}\href
  {\doibase 10.1088/1367-2630/14/10/103035} {\bibfield  {journal} {\bibinfo
  {journal} {New J. Phys.}\ }\textbf {\bibinfo {volume} {14}},\ \bibinfo
  {pages} {103035} (\bibinfo {year} {2012})}\BibitemShut {NoStop}%
\bibitem [{\citenamefont {Liu}\ \emph {et~al.}(2019)\citenamefont {Liu},
  \citenamefont {Song}, \citenamefont {Xue}, \citenamefont {Wang},\ and\
  \citenamefont {Yung}}]{Liu2019Plug}%
  \BibitemOpen
  \bibfield  {author} {\bibinfo {author} {\bibfnamefont {B.-J.}\ \bibnamefont
  {Liu}}, \bibinfo {author} {\bibfnamefont {X.-K.}\ \bibnamefont {Song}},
  \bibinfo {author} {\bibfnamefont {Z.-Y.}\ \bibnamefont {Xue}}, \bibinfo
  {author} {\bibfnamefont {X.}~\bibnamefont {Wang}}, \ and\ \bibinfo {author}
  {\bibfnamefont {M.-H.}\ \bibnamefont {Yung}},\ }\bibfield  {title} {\emph
  {\bibinfo {title} {Plug-and-play approach to nonadiabatic geometric quantum
  gates},\ }}\href {\doibase 10.1103/PhysRevLett.123.100501} {\bibfield
  {journal} {\bibinfo  {journal} {Phys. Rev. Lett.}\ }\textbf {\bibinfo
  {volume} {123}},\ \bibinfo {pages} {100501} (\bibinfo {year}
  {2019})}\BibitemShut {NoStop}%
\bibitem [{\citenamefont {Zheng}\ \emph {et~al.}(2016)\citenamefont {Zheng},
  \citenamefont {Yang},\ and\ \citenamefont {Nori}}]{Zheng2016Comparison}%
  \BibitemOpen
  \bibfield  {author} {\bibinfo {author} {\bibfnamefont {S.-B.}\ \bibnamefont
  {Zheng}}, \bibinfo {author} {\bibfnamefont {C.-P.}\ \bibnamefont {Yang}}, \
  and\ \bibinfo {author} {\bibfnamefont {F.}~\bibnamefont {Nori}},\ }\bibfield
  {title} {\emph {\bibinfo {title} {Comparison of the sensitivity to systematic
  errors between nonadiabatic non-abelian geometric gates and their dynamical
  counterparts},\ }}\href {\doibase 10.1103/PhysRevA.93.032313} {\bibfield
  {journal} {\bibinfo  {journal} {Phys. Rev. A}\ }\textbf {\bibinfo {volume}
  {93}},\ \bibinfo {pages} {032313} (\bibinfo {year} {2016})}\BibitemShut
  {NoStop}%
\bibitem [{\citenamefont {Jing}\ \emph {et~al.}(2017)\citenamefont {Jing},
  \citenamefont {Lam},\ and\ \citenamefont {Wu}}]{Jing2017Non}%
  \BibitemOpen
  \bibfield  {author} {\bibinfo {author} {\bibfnamefont {J.}~\bibnamefont
  {Jing}}, \bibinfo {author} {\bibfnamefont {C.-H.}\ \bibnamefont {Lam}}, \
  and\ \bibinfo {author} {\bibfnamefont {L.-A.}\ \bibnamefont {Wu}},\
  }\bibfield  {title} {\emph {\bibinfo {title} {Non-abelian holonomic
  transformation in the presence of classical noise},\ }}\href {\doibase
  10.1103/PhysRevA.95.012334} {\bibfield  {journal} {\bibinfo  {journal} {Phys.
  Rev. A}\ }\textbf {\bibinfo {volume} {95}},\ \bibinfo {pages} {012334}
  (\bibinfo {year} {2017})}\BibitemShut {NoStop}%
\bibitem [{\citenamefont {Claridge}(2009)}]{Claridge2009high}%
  \BibitemOpen
  \bibfield  {author} {\bibinfo {author} {\bibfnamefont {T.~D.~W.}\
  \bibnamefont {Claridge}},\ }\href@noop {} {\emph {\bibinfo {title}
  {High-Resolution {NMR} Techniques in Organic Chemistry}}}\ (\bibinfo
  {publisher} {Elsevier New York},\ \bibinfo {year} {2009})\BibitemShut
  {NoStop}%
\bibitem [{\citenamefont {Vitanov}\ \emph {et~al.}(2017)\citenamefont
  {Vitanov}, \citenamefont {Rangelov}, \citenamefont {Shore},\ and\
  \citenamefont {Bergmann}}]{Vitanov2017Stiumlated}%
  \BibitemOpen
  \bibfield  {author} {\bibinfo {author} {\bibfnamefont {N.~V.}\ \bibnamefont
  {Vitanov}}, \bibinfo {author} {\bibfnamefont {A.~A.}\ \bibnamefont
  {Rangelov}}, \bibinfo {author} {\bibfnamefont {B.~W.}\ \bibnamefont {Shore}},
  \ and\ \bibinfo {author} {\bibfnamefont {K.}~\bibnamefont {Bergmann}},\
  }\bibfield  {title} {\emph {\bibinfo {title} {Stimulated raman adiabatic
  passage in physics, chemistry, and beyond},\ }}\href {\doibase
  10.1103/RevModPhys.89.015006} {\bibfield  {journal} {\bibinfo  {journal}
  {Rev. Mod. Phys.}\ }\textbf {\bibinfo {volume} {89}},\ \bibinfo {pages}
  {015006} (\bibinfo {year} {2017})}\BibitemShut {NoStop}%
\bibitem [{\citenamefont {Pillet}\ \emph {et~al.}(1993)\citenamefont {Pillet},
  \citenamefont {Valentin}, \citenamefont {Yuan},\ and\ \citenamefont
  {Yu}}]{Pillet1993Adiabatic}%
  \BibitemOpen
  \bibfield  {author} {\bibinfo {author} {\bibfnamefont {P.}~\bibnamefont
  {Pillet}}, \bibinfo {author} {\bibfnamefont {C.}~\bibnamefont {Valentin}},
  \bibinfo {author} {\bibfnamefont {R.-L.}\ \bibnamefont {Yuan}}, \ and\
  \bibinfo {author} {\bibfnamefont {J.}~\bibnamefont {Yu}},\ }\bibfield
  {title} {\emph {\bibinfo {title} {Adiabatic population transfer in a
  multilevel system},\ }}\href {\doibase 10.1103/PhysRevA.48.845} {\bibfield
  {journal} {\bibinfo  {journal} {Phys. Rev. A}\ }\textbf {\bibinfo {volume}
  {48}},\ \bibinfo {pages} {845} (\bibinfo {year} {1993})}\BibitemShut
  {NoStop}%
\bibitem [{\citenamefont {Phillips}(1998)}]{Phillips1998Nobel}%
  \BibitemOpen
  \bibfield  {author} {\bibinfo {author} {\bibfnamefont {W.~D.}\ \bibnamefont
  {Phillips}},\ }\bibfield  {title} {\emph {\bibinfo {title} {Nobel lecture:
  Laser cooling and trapping of neutral atoms},\ }}\href {\doibase
  10.1103/RevModPhys.70.721} {\bibfield  {journal} {\bibinfo  {journal} {Rev.
  Mod. Phys.}\ }\textbf {\bibinfo {volume} {70}},\ \bibinfo {pages} {721}
  (\bibinfo {year} {1998})}\BibitemShut {NoStop}%
\bibitem [{\citenamefont {Garc\'{\i}a-Ripoll}\ and\ \citenamefont
  {Cirac}(2003)}]{Garc2003Quantum}%
  \BibitemOpen
  \bibfield  {author} {\bibinfo {author} {\bibfnamefont {J.~J.}\ \bibnamefont
  {Garc\'{\i}a-Ripoll}}\ and\ \bibinfo {author} {\bibfnamefont {J.~I.}\
  \bibnamefont {Cirac}},\ }\bibfield  {title} {\emph {\bibinfo {title} {Quantum
  computation with unknown parameters},\ }}\href {\doibase
  10.1103/PhysRevLett.90.127902} {\bibfield  {journal} {\bibinfo  {journal}
  {Phys. Rev. Lett.}\ }\textbf {\bibinfo {volume} {90}},\ \bibinfo {pages}
  {127902} (\bibinfo {year} {2003})}\BibitemShut {NoStop}%
\bibitem [{\citenamefont {Daems}\ and\ \citenamefont
  {Gu\'erin}(2007)}]{Daems2007Adiabatic}%
  \BibitemOpen
  \bibfield  {author} {\bibinfo {author} {\bibfnamefont {D.}~\bibnamefont
  {Daems}}\ and\ \bibinfo {author} {\bibfnamefont {S.}~\bibnamefont
  {Gu\'erin}},\ }\bibfield  {title} {\emph {\bibinfo {title} {Adiabatic quantum
  search scheme with atoms in a cavity driven by lasers},\ }}\href {\doibase
  10.1103/PhysRevLett.99.170503} {\bibfield  {journal} {\bibinfo  {journal}
  {Phys. Rev. Lett.}\ }\textbf {\bibinfo {volume} {99}},\ \bibinfo {pages}
  {170503} (\bibinfo {year} {2007})}\BibitemShut {NoStop}%
\bibitem [{\citenamefont {Jing}\ \emph {et~al.}(2016)\citenamefont {Jing},
  \citenamefont {Sarandy}, \citenamefont {Lidar}, \citenamefont {Luo},\ and\
  \citenamefont {Wu}}]{Jing2016Eigenstate}%
  \BibitemOpen
  \bibfield  {author} {\bibinfo {author} {\bibfnamefont {J.}~\bibnamefont
  {Jing}}, \bibinfo {author} {\bibfnamefont {M.~S.}\ \bibnamefont {Sarandy}},
  \bibinfo {author} {\bibfnamefont {D.~A.}\ \bibnamefont {Lidar}}, \bibinfo
  {author} {\bibfnamefont {D.-W.}\ \bibnamefont {Luo}}, \ and\ \bibinfo
  {author} {\bibfnamefont {L.-A.}\ \bibnamefont {Wu}},\ }\bibfield  {title}
  {\emph {\bibinfo {title} {Eigenstate tracking in open quantum systems},\
  }}\href {\doibase 10.1103/PhysRevA.94.042131} {\bibfield  {journal} {\bibinfo
   {journal} {Phys. Rev. A}\ }\textbf {\bibinfo {volume} {94}},\ \bibinfo
  {pages} {042131} (\bibinfo {year} {2016})}\BibitemShut {NoStop}%
\bibitem [{\citenamefont {Chen}\ \emph
  {et~al.}(2010{\natexlab{a}})\citenamefont {Chen}, \citenamefont {Lizuain},
  \citenamefont {Ruschhaupt}, \citenamefont {Gu\'ery-Odelin},\ and\
  \citenamefont {Muga}}]{Chen2010Shortcut}%
  \BibitemOpen
  \bibfield  {author} {\bibinfo {author} {\bibfnamefont {X.}~\bibnamefont
  {Chen}}, \bibinfo {author} {\bibfnamefont {I.}~\bibnamefont {Lizuain}},
  \bibinfo {author} {\bibfnamefont {A.}~\bibnamefont {Ruschhaupt}}, \bibinfo
  {author} {\bibfnamefont {D.}~\bibnamefont {Gu\'ery-Odelin}}, \ and\ \bibinfo
  {author} {\bibfnamefont {J.~G.}\ \bibnamefont {Muga}},\ }\bibfield  {title}
  {\emph {\bibinfo {title} {Shortcut to adiabatic passage in two- and
  three-level atoms},\ }}\href {\doibase 10.1103/PhysRevLett.105.123003}
  {\bibfield  {journal} {\bibinfo  {journal} {Phys. Rev. Lett.}\ }\textbf
  {\bibinfo {volume} {105}},\ \bibinfo {pages} {123003} (\bibinfo {year}
  {2010}{\natexlab{a}})}\BibitemShut {NoStop}%
\bibitem [{\citenamefont {Gu\'ery-Odelin}\ \emph {et~al.}(2019)\citenamefont
  {Gu\'ery-Odelin}, \citenamefont {Ruschhaupt}, \citenamefont {Kiely},
  \citenamefont {Torrontegui}, \citenamefont {Mart\'{\i}nez-Garaot},\ and\
  \citenamefont {Muga}}]{Guery2019Shortcuts}%
  \BibitemOpen
  \bibfield  {author} {\bibinfo {author} {\bibfnamefont {D.}~\bibnamefont
  {Gu\'ery-Odelin}}, \bibinfo {author} {\bibfnamefont {A.}~\bibnamefont
  {Ruschhaupt}}, \bibinfo {author} {\bibfnamefont {A.}~\bibnamefont {Kiely}},
  \bibinfo {author} {\bibfnamefont {E.}~\bibnamefont {Torrontegui}}, \bibinfo
  {author} {\bibfnamefont {S.}~\bibnamefont {Mart\'{\i}nez-Garaot}}, \ and\
  \bibinfo {author} {\bibfnamefont {J.~G.}\ \bibnamefont {Muga}},\ }\bibfield
  {title} {\emph {\bibinfo {title} {Shortcuts to adiabaticity: Concepts,
  methods, and applications},\ }}\href {\doibase 10.1103/RevModPhys.91.045001}
  {\bibfield  {journal} {\bibinfo  {journal} {Rev. Mod. Phys.}\ }\textbf
  {\bibinfo {volume} {91}},\ \bibinfo {pages} {045001} (\bibinfo {year}
  {2019})}\BibitemShut {NoStop}%
\bibitem [{\citenamefont {Baksic}\ \emph {et~al.}(2016)\citenamefont {Baksic},
  \citenamefont {Ribeiro},\ and\ \citenamefont {Clerk}}]{Baksic2016Speeding}%
  \BibitemOpen
  \bibfield  {author} {\bibinfo {author} {\bibfnamefont {A.}~\bibnamefont
  {Baksic}}, \bibinfo {author} {\bibfnamefont {H.}~\bibnamefont {Ribeiro}}, \
  and\ \bibinfo {author} {\bibfnamefont {A.~A.}\ \bibnamefont {Clerk}},\
  }\bibfield  {title} {\emph {\bibinfo {title} {Speeding up adiabatic quantum
  state transfer by using dressed states},\ }}\href {\doibase
  10.1103/PhysRevLett.116.230503} {\bibfield  {journal} {\bibinfo  {journal}
  {Phys. Rev. Lett.}\ }\textbf {\bibinfo {volume} {116}},\ \bibinfo {pages}
  {230503} (\bibinfo {year} {2016})}\BibitemShut {NoStop}%
\bibitem [{\citenamefont {Li}\ and\ \citenamefont
  {Chen}(2016)}]{Li2016Shortcut}%
  \BibitemOpen
  \bibfield  {author} {\bibinfo {author} {\bibfnamefont {Y.-C.}\ \bibnamefont
  {Li}}\ and\ \bibinfo {author} {\bibfnamefont {X.}~\bibnamefont {Chen}},\
  }\bibfield  {title} {\emph {\bibinfo {title} {Shortcut to adiabatic
  population transfer in quantum three-level systems: Effective two-level
  problems and feasible counterdiabatic driving},\ }}\href {\doibase
  10.1103/PhysRevA.94.063411} {\bibfield  {journal} {\bibinfo  {journal} {Phys.
  Rev. A}\ }\textbf {\bibinfo {volume} {94}},\ \bibinfo {pages} {063411}
  (\bibinfo {year} {2016})}\BibitemShut {NoStop}%
\bibitem [{\citenamefont {Chen}\ \emph
  {et~al.}(2010{\natexlab{b}})\citenamefont {Chen}, \citenamefont {Ruschhaupt},
  \citenamefont {Schmidt}, \citenamefont {del Campo}, \citenamefont
  {Gu\'ery-Odelin},\ and\ \citenamefont {Muga}}]{Chen2010Fast}%
  \BibitemOpen
  \bibfield  {author} {\bibinfo {author} {\bibfnamefont {X.}~\bibnamefont
  {Chen}}, \bibinfo {author} {\bibfnamefont {A.}~\bibnamefont {Ruschhaupt}},
  \bibinfo {author} {\bibfnamefont {S.}~\bibnamefont {Schmidt}}, \bibinfo
  {author} {\bibfnamefont {A.}~\bibnamefont {del Campo}}, \bibinfo {author}
  {\bibfnamefont {D.}~\bibnamefont {Gu\'ery-Odelin}}, \ and\ \bibinfo {author}
  {\bibfnamefont {J.~G.}\ \bibnamefont {Muga}},\ }\bibfield  {title} {\emph
  {\bibinfo {title} {Fast optimal frictionless atom cooling in harmonic traps:
  Shortcut to adiabaticity},\ }}\href {\doibase 10.1103/PhysRevLett.104.063002}
  {\bibfield  {journal} {\bibinfo  {journal} {Phys. Rev. Lett.}\ }\textbf
  {\bibinfo {volume} {104}},\ \bibinfo {pages} {063002} (\bibinfo {year}
  {2010}{\natexlab{b}})}\BibitemShut {NoStop}%
\bibitem [{\citenamefont {Chen}\ \emph {et~al.}(2011)\citenamefont {Chen},
  \citenamefont {Torrontegui},\ and\ \citenamefont {Muga}}]{Chen2011Lewis}%
  \BibitemOpen
  \bibfield  {author} {\bibinfo {author} {\bibfnamefont {X.}~\bibnamefont
  {Chen}}, \bibinfo {author} {\bibfnamefont {E.}~\bibnamefont {Torrontegui}}, \
  and\ \bibinfo {author} {\bibfnamefont {J.~G.}\ \bibnamefont {Muga}},\
  }\bibfield  {title} {\emph {\bibinfo {title} {Lewis-riesenfeld invariants and
  transitionless quantum driving},\ }}\href {\doibase
  10.1103/PhysRevA.83.062116} {\bibfield  {journal} {\bibinfo  {journal} {Phys.
  Rev. A}\ }\textbf {\bibinfo {volume} {83}},\ \bibinfo {pages} {062116}
  (\bibinfo {year} {2011})}\BibitemShut {NoStop}%
\bibitem [{\citenamefont {Chen}\ and\ \citenamefont
  {Muga}(2012)}]{Chen2012Engineering}%
  \BibitemOpen
  \bibfield  {author} {\bibinfo {author} {\bibfnamefont {X.}~\bibnamefont
  {Chen}}\ and\ \bibinfo {author} {\bibfnamefont {J.~G.}\ \bibnamefont
  {Muga}},\ }\bibfield  {title} {\emph {\bibinfo {title} {Engineering of fast
  population transfer in three-level systems},\ }}\href {\doibase
  10.1103/PhysRevA.86.033405} {\bibfield  {journal} {\bibinfo  {journal} {Phys.
  Rev. A}\ }\textbf {\bibinfo {volume} {86}},\ \bibinfo {pages} {033405}
  (\bibinfo {year} {2012})}\BibitemShut {NoStop}%
\bibitem [{\citenamefont {Qi}\ and\ \citenamefont
  {Jing}(2022)}]{Qi2022Accelerated}%
  \BibitemOpen
  \bibfield  {author} {\bibinfo {author} {\bibfnamefont {S.-F.}\ \bibnamefont
  {Qi}}\ and\ \bibinfo {author} {\bibfnamefont {J.}~\bibnamefont {Jing}},\
  }\bibfield  {title} {\emph {\bibinfo {title} {Accelerated adiabatic passage
  in cavity magnomechanics},\ }}\href {\doibase 10.1103/PhysRevA.105.053710}
  {\bibfield  {journal} {\bibinfo  {journal} {Phys. Rev. A}\ }\textbf {\bibinfo
  {volume} {105}},\ \bibinfo {pages} {053710} (\bibinfo {year}
  {2022})}\BibitemShut {NoStop}%
\bibitem [{\citenamefont {An}\ \emph {et~al.}(2016)\citenamefont {An},
  \citenamefont {Lv}, \citenamefont {del Campo},\ and\ \citenamefont
  {Kim}}]{An2016Shortcuts}%
  \BibitemOpen
  \bibfield  {author} {\bibinfo {author} {\bibfnamefont {S.}~\bibnamefont
  {An}}, \bibinfo {author} {\bibfnamefont {D.}~\bibnamefont {Lv}}, \bibinfo
  {author} {\bibfnamefont {A.}~\bibnamefont {del Campo}}, \ and\ \bibinfo
  {author} {\bibfnamefont {K.}~\bibnamefont {Kim}},\ }\bibfield  {title} {\emph
  {\bibinfo {title} {Shortcuts to adiabaticity by counterdiabatic driving for
  trapped-ion displacement in phase space},\ }}\href {\doibase
  10.1038/ncomms12999} {\bibfield  {journal} {\bibinfo  {journal} {Nat.
  Commun.}\ }\textbf {\bibinfo {volume} {7}},\ \bibinfo {pages} {12999}
  (\bibinfo {year} {2016})}\BibitemShut {NoStop}%
\bibitem [{\citenamefont {Berry}(2009)}]{Berry2009Transition}%
  \BibitemOpen
  \bibfield  {author} {\bibinfo {author} {\bibfnamefont {M.~V.}\ \bibnamefont
  {Berry}},\ }\bibfield  {title} {\emph {\bibinfo {title} {Transitionless
  quantum driving},\ }}\href {\doibase 10.1088/1751-8113/42/36/365303}
  {\bibfield  {journal} {\bibinfo  {journal} {J. Phys. A}\ }\textbf {\bibinfo
  {volume} {42}},\ \bibinfo {pages} {365303} (\bibinfo {year}
  {2009})}\BibitemShut {NoStop}%
\bibitem [{\citenamefont {Koch}\ \emph {et~al.}(2007)\citenamefont {Koch},
  \citenamefont {Yu}, \citenamefont {Gambetta}, \citenamefont {Houck},
  \citenamefont {Schuster}, \citenamefont {Majer}, \citenamefont {Blais},
  \citenamefont {Devoret}, \citenamefont {Girvin},\ and\ \citenamefont
  {Schoelkopf}}]{Koch2007Charge}%
  \BibitemOpen
  \bibfield  {author} {\bibinfo {author} {\bibfnamefont {J.}~\bibnamefont
  {Koch}}, \bibinfo {author} {\bibfnamefont {T.~M.}\ \bibnamefont {Yu}},
  \bibinfo {author} {\bibfnamefont {J.}~\bibnamefont {Gambetta}}, \bibinfo
  {author} {\bibfnamefont {A.~A.}\ \bibnamefont {Houck}}, \bibinfo {author}
  {\bibfnamefont {D.~I.}\ \bibnamefont {Schuster}}, \bibinfo {author}
  {\bibfnamefont {J.}~\bibnamefont {Majer}}, \bibinfo {author} {\bibfnamefont
  {A.}~\bibnamefont {Blais}}, \bibinfo {author} {\bibfnamefont {M.~H.}\
  \bibnamefont {Devoret}}, \bibinfo {author} {\bibfnamefont {S.~M.}\
  \bibnamefont {Girvin}}, \ and\ \bibinfo {author} {\bibfnamefont {R.~J.}\
  \bibnamefont {Schoelkopf}},\ }\bibfield  {title} {\emph {\bibinfo {title}
  {Charge-insensitive qubit design derived from the cooper pair box},\ }}\href
  {\doibase 10.1103/PhysRevA.76.042319} {\bibfield  {journal} {\bibinfo
  {journal} {Phys. Rev. A}\ }\textbf {\bibinfo {volume} {76}},\ \bibinfo
  {pages} {042319} (\bibinfo {year} {2007})}\BibitemShut {NoStop}%
\bibitem [{\citenamefont {Vepsäläinen}\ \emph {et~al.}(2019)\citenamefont
  {Vepsäläinen}, \citenamefont {Danilin},\ and\ \citenamefont
  {Paraoanu}}]{Antti2019Superadiabatic}%
  \BibitemOpen
  \bibfield  {author} {\bibinfo {author} {\bibfnamefont {A.}~\bibnamefont
  {Vepsäläinen}}, \bibinfo {author} {\bibfnamefont {S.}~\bibnamefont
  {Danilin}}, \ and\ \bibinfo {author} {\bibfnamefont {G.~S.}\ \bibnamefont
  {Paraoanu}},\ }\bibfield  {title} {\emph {\bibinfo {title} {Superadiabatic
  population transfer in a three-level superconducting circuit},\ }}\href
  {\doibase 10.1126/sciadv.aau5999} {\bibfield  {journal} {\bibinfo  {journal}
  {Sci. Adv.}\ }\textbf {\bibinfo {volume} {5}},\ \bibinfo {pages} {eaau5999}
  (\bibinfo {year} {2019})}\BibitemShut {NoStop}%
\bibitem [{\citenamefont {Ban}\ \emph {et~al.}(2018)\citenamefont {Ban},
  \citenamefont {Jiang}, \citenamefont {Li}, \citenamefont {Wang},\ and\
  \citenamefont {Chen}}]{Yue2018Fast}%
  \BibitemOpen
  \bibfield  {author} {\bibinfo {author} {\bibfnamefont {Y.}~\bibnamefont
  {Ban}}, \bibinfo {author} {\bibfnamefont {L.-X.}\ \bibnamefont {Jiang}},
  \bibinfo {author} {\bibfnamefont {Y.-C.}\ \bibnamefont {Li}}, \bibinfo
  {author} {\bibfnamefont {L.-J.}\ \bibnamefont {Wang}}, \ and\ \bibinfo
  {author} {\bibfnamefont {X.}~\bibnamefont {Chen}},\ }\bibfield  {title}
  {\emph {\bibinfo {title} {Fast creation and transfer of coherence in triple
  quantum dots by using shortcuts to adiabaticity},\ }}\href {\doibase
  10.1364/OE.26.031137} {\bibfield  {journal} {\bibinfo  {journal} {Opt.
  Express}\ }\textbf {\bibinfo {volume} {26}},\ \bibinfo {pages} {31137}
  (\bibinfo {year} {2018})}\BibitemShut {NoStop}%
\end{thebibliography}%
\end{document}